\documentclass[journal]{IEEEtran}

\usepackage{bbm}
\usepackage{cite}
\usepackage{amsmath}
\usepackage{amssymb}
\usepackage{amsfonts}
\usepackage{graphicx}
\usepackage{epstopdf}
\usepackage{algorithm}
\usepackage{algorithmic}
\usepackage{epsfig}
\usepackage{epstopdf}
\usepackage{subfigure}
\usepackage{url}

\usepackage{hyperref}
\hypersetup{
	colorlinks=true,
	linkcolor=blue,
	citecolor=blue,
	urlcolor=blue 
}

\usepackage{pifont}
\usepackage{psfrag}
\usepackage{times}
\usepackage{booktabs}
\usepackage[T1]{fontenc}
\usepackage{moreverb}
\usepackage{color}
\usepackage{setspace}
\usepackage{mathrsfs}
\usepackage{array}
\usepackage{txfonts}
\usepackage{makecell}
\usepackage{bm}

\begin{document}
	\title{Bringing Environmental Enhancement Back to Its Physical Essence via Specular Reflecting Surfaces
		\author{Qingxiao Huang,~\IEEEmembership{Graduate Student Member,~IEEE}, Qianyao Ren, Yiqin Deng,~\IEEEmembership{Member,~IEEE},\\ Jun Huang,~\IEEEmembership{Member,~IEEE}, Yang Kun,~\IEEEmembership{Fellow,~IEEE} and Yuguang Fang,~\IEEEmembership{Fellow,~IEEE}
			\thanks{The research work described in this paper was conducted in the JC STEM Lab of Smart City funded by The Hong Kong Jockey Club Charities Trust under Contract 2023-0108. This work was supported in part by a grant from the Research Grants Council of the Hong Kong Special Administrative Region, China (Project No. CityU 11216324) and in part by the Hong Kong SAR Government under the Global STEM Professorship and Research Talent Hub. This paper will be presented in part at the IEEE GLOBECOM’26.}
			\thanks{Qingxiao Huang, Qianyao Ren, Jun Huang and Yuguang Fang are with the Hong Kong Jockey Club STEM Lab of Smart City and the Department of Computer Science, City University of Hong Kong, Hong Kong, e-mail: qx.huang@my.cityu.edu.hk, qianyaren2-c@my.cityu.edu.hk, jun.huang@cityu.edu.hk and my.Fang@cityu.edu.hk.}
			\thanks{Yiqin Deng is with the School of Data Science, Lingnan University, Hong Kong, e-mail: yiqindeng@ln.edu.hk.}
\thanks{Kun Yang is with the State Key Laboratory of Novel Software Technology, Nanjing University, Nanjing, 210008, China, Institute of Intelligent Networks and Communications (NINE), Nanjing University (Suzhou Campus), Suzhou, 215163, China, e-mail: kunyang@nju.edu.cn.}
		}
	}
	\maketitle
	\begin{abstract}
Intelligent control of wireless propagation environments is crucial for future network capacity and reliability. Unlike circuit-controlled reconfigurable intelligent surfaces (RIS), mechanically actuated specular reflecting surfaces (SRS) offer a simpler and potentially more cost-effective alternative. In this paper, based on the tractable ray-based cascaded channel model with power-projection correction, we investigate the fundamental operational behaviors of an ideal SRS in free space. Specifically, in the angle-aligned near field, edge reflections cause non-constructive combining, resulting in a damped oscillatory convergence of the gain to an aperture-independent constant. We further obtain the far-field behavior, unbounded-aperture asymptotics, an optimal aperture size and reflection angle, and a gain-based near/far-field boundary. For misalignment, we provide accurate approximations for small and large apertures via center-point and stationary-point analyses. We also define the SRS beam pattern, derive analytical 3-dB beamwidths, and quantify the effective region where a main lobe exists. Finally, we derive a closed-form achievable-rate for an SRS-aided communication system. Numerical results validate the proposed expressions, reveal distinct near-/far-field behaviors of specular reflection, and show that SRS can outperform RIS in the far field due to continuous aperture and angular-resolution control and stronger power projection.

	\end{abstract}
	\begin{IEEEkeywords}
		Specular reflecting surface (SRS), reconfigurable intelligent surface (RIS), near-field communication, environmental enhancement.
	\end{IEEEkeywords}

\section{Introduction}
\subsection{Background}
Smart cities are evolving into highly interconnected cyber--physical ecosystems, where large-scale Internet-of-Things (IoT) deployments, real-time digital twins, immersive extended reality (XR), and autonomous transportation demand massive data-exchange \cite{10697414}. Meanwhile, dense deployments and complex urban propagation conditions, such as severe blockage and strong interference, make it increasingly difficult to sustain stable links. Collectively, these trends call for next-generation wireless networks that must not only deliver higher capacity but also satisfy stringent and diverse quality-of-service (QoS) requirements, including low latency, high reliability, and massive connectivity \cite{10697414}.

Since the fifth-generation (5G) wireless networks, massive multiple-input multiple-output (MIMO) has become a key enabling technology. Looking towards 6G, even larger-scale arrays are expected to be deployed, and the resulting paradigm shift in propagation physics, namely spherical-wave channels in the near field, will be further exploited to improve spectral efficiency \cite{10558818}. In addition, a variety of aggressive transceiver architectures have emerged, such as holographic MIMO \cite{9136592, 11433651} and flexible intelligent metasurfaces (FIM) \cite{10850658}. Beyond optimizing transmitters and receivers, enhancing and controlling the wireless propagation environment \cite{9082859} is also essential for further spectral-efficiency gains, particularly in dense urban scenarios where coverage holes and blockage are prevalent.

\subsection{Related Works}
Reconfigurable intelligent surfaces (RIS) have emerged as a promising paradigm for building smart radio environments \cite{8910627}. An RIS typically consists of a large number of sub-wavelength reflecting elements whose electromagnetic responses can be tuned by control circuits \cite{9140329}. By adjusting the phase shifts (and possibly amplitudes) across the surface, RIS can reshape the impinging wavefronts and steer the reflected signals toward intended directions, thereby enhancing coverage and improving spectral efficiency without deploying additional active radio-frequency chains \cite{zhang2021reconfigurable}. 

Although increasing the number of RIS elements can provide higher spatial gains and enable simplified beamforming designs thanks to the extremely high spatial resolution \cite{10559446,10470405}, it also poses significant challenges for practical deployment. In particular, RIS is inherently passive and cannot perform channel estimation by itself, making channel acquisition dependent on additional active elements or base stations \cite{Liang2021RISSmartWirelessEnvironments}. The resulting channel-estimation overhead can be prohibitive, especially for large-scale RIS deployments or multi-hop RIS networks \cite{ma2024multi}. Moreover, due to circuit coupling among closely spaced unit cells and impedance mismatches \cite{9140329}, it is difficult for practical RIS elements to achieve high-precision, high-resolution discrete phase control. As a result, compared with certain relay or signal-boosting devices, RIS does not necessarily offer a clear cost advantage. These limitations motivate alternative approaches that can retain the key benefit of environmental enhancement while substantially simplifying hardware and control.

In early studies, parabolic reflectors were used to enhance the received signals of radio telescopes by exploiting geometric focusing \cite{1687092}. Reflecting surfaces were also advocated as passive satellite reflectors to enable long-distance communication links with low complexity and cost \cite{4066064}. This idea was echoed by Project Echo, which employed metallized balloon satellites as passive reflectors to conduct long-range communication experiments and measure link losses \cite{Jakes1961EchoI}. In millimeter-wave (mmWave) wireless, reflection-based mechanisms have also been widely exploited \cite{7914640,10.1145/2342356.2342440,9384308,9374714}. Xue \emph{et al.} \cite{7914640} investigated a single-user mmWave communication system, where environmental reflections from walls and floors were leveraged to enable multi-beam transmission. Besides, Zhou \emph{et al.} \cite{10.1145/2342356.2342440} enabled mmWave communications among multiple transceiver pairs by deploying metallic reflectors on the data-center ceiling, and validated the effectiveness of the proposed design using a hardware testbed. Moreover, Li \emph{et al.} \cite{9384308} investigated a multi-target localization system around corners, where a non-line-of-sight (NLOS) localization algorithm was designed by exploiting multiple specular reflections from walls. Furthermore, Anjinappa \emph{et al.} \cite{9374714} studied an mmWave base-station communication system, where a joint deployment strategy for base stations and reflecting panels was developed to expand network coverage and reduce deployment cost. 

To further increase the controllability, reflecting surfaces can be mounted on motor-driven rotational shafts to enable rotation control, or placed on sliding platforms to enable position control. Such mechatronic actuation techniques have already been widely adopted in transceiver architectures, e.g., in movable antenna systems\cite{11007274, zheng2025}. Specifically, Lu \emph{et al.} \cite{10279522} established an electromagnetic theory based far-field channel model for specular reflecting surface (SRS) in free-space and then validated it via real-world measurements. Furthermore, based on the channel model, they studied an SRS-aided communication system in the sub-6~GHz band, where both translation- and rotation-based actuation were considered \cite{10989638}. They proposed a joint placement-and-rotation optimization algorithm to maximize the minimum received power over the target coverage regions of multiple SRSs. Moreover, Xu \emph{et al.} \cite{11082321} investigated an SRS-aided visible light communication system, where ray tracing was employed to model indoor light propagation, and a rotation-optimization algorithm was proposed to increase the received optical power. A prototype was also built to experimentally validate the proposed approach. Besides, Zheng \emph{et al.} \cite{zheng2025sensing} extended SRS to transmissive surfaces and studied a mechanically actuated double-layer transmissive-surface-aided wireless sensing system, where a mechanical beamforming algorithm was proposed to maximize the minimum SINR of the echo signals. Our prior work \cite{9013979,zhang2019tunable} investigated a mechanically actuated SRS-assisted mmWave vehicular communication system, where a learning-based scheme was proposed to reshape the propagation environment. Specifically, the SRS reflection angle was adaptively adjusted based on observable traffic patterns to improve link quality under dynamic mobility and blockage. We further extended this line of research to multi-SRS-assisted mmWave WLANs \cite{9082859}, where the system architecture, operational procedures, and signaling process were developed to enhance coverage and reliability. We also studied a sub-6~GHz vehicular network enhanced by continuously deployed roadside SRSs, where a dynamic rotation strategy was designed for a group of virtual SRSs to enable large-area signal enhancement and interference management \cite{ren2026}. The proposed schemes can be transparently integrated into existing communication systems, and the channel rank is significantly improved with the aid of SRSs.

\subsection{Motivation and Contributions}

However, existing researches have the following drawbacks.
\begin{itemize}
	\item Most current SRS-based radio frequency systems are developed under a far-field channel assumption, while the near-/far-field behavior of a single SRS has not been sufficiently investigated. For an SRS that cannot control the signal over every point on the surface, reflections from the edge aperture may be combined in a non-costructive fashion in the near field. 
	\item Over a sufficiently large service region, a single large-aperture SRS may already be adequate to establish a new line-of-sight (LOS) link, making an array of multiple small SRSs unnecessary. 
	\item Although various optimization algorithms have been proposed for SRS-aided communication and sensing systems, several fundamental properties of single-SRS reflection remain underexplored, such as the near-/far-field regimes, beam pattern, beamwidth, and effective operating region. 
	\item While many works highlight the ultra-low-cost advantage of natural reflections from metallic panels over circuit-based RIS, other potential benefits have yet to be demonstrated, including the advantages of a continuous reflecting surface and the higher control resolution enabled by continuous mechanical rotation.
\end{itemize}
Therefore, this work aims to explore the fundamental properties of a single SRS, with a particular focus on near-field effects, to facilitate more advanced and diversified SRS research in the future. In light of these, our novel contributions are summarized as follows:
\begin{itemize}
	\item Based on a ray-based channel model with power-projection correction, we derive closed-form expressions for the cascaded near-field channel and gain of an aligned SRS link, and further analyze their far-field behavior and the asymptotics as the aperture size tends to infinity. In addition, we obtain the optimal aperture size and reflection angle, as well as the SRS near-/far-field boundary defined in terms of channel gain.
	
	\item Furthermore, we analyze the general misaligned case where the reflection angle is not equal to the angle of incidence, and derive closed-form expressions for the channel and gain in both the near- and far-field regimes. By exploiting the stationary point of the phase, we characterize the asymptotic behavior for finite apertures and provide analytical expressions for SRS with large apertures.
	
	\item In addition, we define the SRS beam pattern and derive the 3-dB beamwidth in the far field, the beamwidth equation in the near field, and the effective operating region of SRS in which a main lobe exists. Moreover, building on previous results, we establish an SRS-aided wireless communication system model and obtain a closed-form solution for the optimal rotation angle.
	
	\item Simulation results validate the accuracy of the derived closed-form expressions, reveal the distinct near- and far-field characteristics and effective operating range of specular SRS reflection, and demonstrate that, compared with RIS, SRS offers multiple advantages in forming a new NLOS multipath component with two LOS hops (the so-called cascade channel), including a continuous aperture, continuous angular control, and stronger power projection.
\end{itemize}

The rest of this paper is organized as follows. In Section II, we investigate the angle-aligned channel and the near-/far-field boundary, while the angle-misaligned channel as well as the beam pattern and beamwidth are studied in Section~III. The SRS-aided communication system is examined in Section~IV. Then, our numerical results are presented in Section V, followed by the conclusions in Section VI.

Notations: $|a|$ and $||\mathbf{a}||$ are the magnitude and norm of a scalar $a$ and vector $\mathbf{a}$, respectively; $||\mathbf{A}||$ denotes the Frobenius norm of the matrix $\mathbf{A}$; $\mathbf{A}(i,j)$ represents the specific element in the $i$-th row and $j$-th column of $\mathbf{A}$; $\mathbf{a}(i)$ denotes the $i$-th element of the vector $\mathbf{a}$; ``:='' is signified as ``is defined as''.

\section{Analysis of SRS: the Angle-aligned Case}
\subsection{Overview of SRS}	
Unlike RIS, SRS dispenses with the expensive circuit-based phase-control architecture. Instead, it aims to employ low-cost materials, such as metallic panels that have been widely considered in existing studies. Moreover, in contrast to RIS unit cells, SRS does not suffer from circuit-coupling effects among half-wavelength-spaced elements. A high reflection coefficient and a wide reflection-angle range are key criteria for material selection. Beyond rigid panels, flexible surfaces as well as liquid-, gas-, or hybrid-material implementations remain promising directions for further exploration. In this paper, we focus on a smooth reflecting surface with rotation-based control.

Moreover, although mechanical actuation is typically slower than electronic control, many slowly varying tasks, such as coverage enhancement, region-of-interest signal boosting, and periodic environmental monitoring, do not strictly require sub-millisecond reconfiguration agility. In these settings, beam updates are typically required only on the millisecond-to-second timescale, which is well within the capability of mechanical reconfiguration.

Furthermore, SRS focuses on forming a new NLOS multipath component with two LOS hops between the transmitter and receiver to enhance the end-to-end channel gain and degrees of freedom. Accordingly, channel estimation can sometimes be simplified to estimating the LOS links between the SRS and the transceivers, where emerging application-layer sensing techniques, such as vision-based sensing or millimeter-wave radar, can be leveraged. When SRS is employed for coverage enhancement or signal boosting, additional SRS-specific cascaded channel estimation may not be necessary. Instead, existing base-station--to--user channel estimates can be reused, and the SRS naturally appears to be a dominant multipath component with a large eigenvalue in the channel.

Finally, although electromagnetic theory based analyses can further reveal polarization-related effects, we focus on fundamental beam behaviors, especially near-field effects. Therefore, we adopt a simpler and more intuitive ray-based model in the sequel, which has long been widely used as a tractable abstraction in wireless communication analysis.
\subsection{Cascaded Channel and Channel Gain}
\vspace{-4pt}
\begin{figure}[h]
	\centering
	\includegraphics[width=70mm]{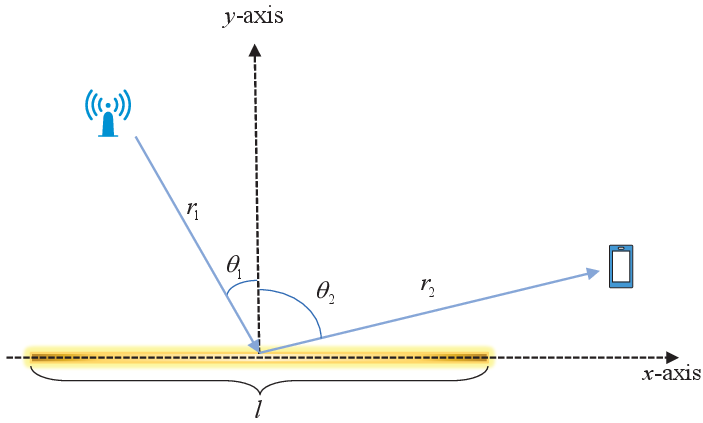}\\
	\caption{Planar schematic of the SRS-aided communication system} 
	\label{Single_SRS}
	\vspace{-4pt}
\end{figure}
SRS aims to align the transmitter and receiver via specular reflection by rotating the entire surface, which means the angle of incidence from the transmitter to the SRS center equals the angle of reflection from the SRS center to the receiver. Therefore, we first analyze the channel and gain of SRS under the angle-aligned configuration. Without loss of generality, we conduct analysis under a coplanar setting: all nodes lie in the same \(x\!-\!o-\!y\) plane, and the structure is assumed to have unit extent along \(z\) axis. As illustrated in Fig.~1, we establish a Cartesian coordinate system with the SR center as the origin and the  SRS coincident with the \(x\)-axis. The width of the \(\mathrm{SRS}\) is \(l\). Consider an arbitrary point in space and an arbitrary point $(x,0)$ on the SRS, the angle and distance between them are denoted by $\theta$ and $r$, respectively, and the channel between them is expressed as \cite{9866003,9184098}
\begin{align}
	h(r,\theta,x)=\sqrt{\alpha(r,\theta,x)}e^{j k \sqrt{(r\sin\theta-x)^2+(r\cos\theta)^2}}
\end{align}	
where
\begin{align}
	\alpha(r,\theta,x)=\underbrace{\frac{1}{4\pi\left( (r\sin\theta-x)^2+(r\cos\theta)^2\right) }}_{\text{Free-space propagation}}\underbrace{\frac{r\cos\theta}{\sqrt{ (r\sin\theta-x)^2+(r\cos\theta)^2}}}_{\text{Effective power projection}},
\end{align}	
where $k=\frac{2\pi}{\lambda}$ and $\lambda$ is the wavelength. The first part of $\alpha(r,\theta,x)$ accounts for the free-space propagation loss, while the second part represents the effective power projection induced by the placement of the SRS.
 
Let the angle of incidence from the transmitter to the SR center and the angle of reflection from the SR center to the receiver both be \(\theta\), and denote the corresponding distances by \(r_1\) and \(r_2\), respectively. Assuming an ideal specular reflection, with a reflection coefficient of \(1\) and no diffuse scattering, the cascaded channel between the transmitter and receiver via the SRS reflection can be expressed as

\begin{align}\label{eq:h_def}
h(r_1,r_2,\theta)=&\int_{-l/2}^{l/2}\frac{\sqrt{r_1 r_2}\cos\theta}{4\pi\left( (r_1\sin\theta+x)^2+(r_1\cos\theta)^2\right)^{\frac{3}{4}} }\nonumber\\ &\times\frac{1}{\left( (r_2\sin\theta-x)^2+(r_2\cos\theta)^2\right)^{\frac{3}{4}}}\nonumber\\ &\times e^{ j k(\sqrt{(r_1\sin\theta+x)^2+(r_1\cos\theta)^2}+\sqrt{(r_2\sin\theta-x)^2+(r_2\cos\theta)^2})}  \mathrm{d}x
\nonumber\\ =&\int_{-l/2}^{l/2} A(x) \exp\left\lbrace jk \left(d_1(x)+ d_2(x) \right) \right\rbrace   \mathrm{d}x
\end{align}	
where $A(x)\!:=\!\! \frac{\sqrt{r_1 r_2}\cos\theta}{4\pi d_2^{3/2}(x)d_1^{3/2}(x)}$, $d_1(x)\!:=\!\! \sqrt{(r_1\sin\theta+x)^2+(r_1\cos\theta)^2}$ and $d_2(x):= \sqrt{(r_2\sin\theta-x)^2+(r_2\cos\theta)^2}$.

In general, to achieve a better specular reflection effect, both the transmitter and receiver are preferably located in the far-field region of the SRS. Since electromagnetic waves can be approximated as plane waves in the far field, the phase difference of the incident and reflected channels across different points on the continuous surface depends only on the angle. When the angle of incidence equals the angle of reflection, the phases at all points become identical, thus the signals add constructively. Moreover, since \(l \ll r_1, r_2\), the amplitude factor \(A(x)\) can be approximated by its value \(A(0)= \frac{\cos\theta}{4\pi r_1 r_2}\) at the center. In this case, the reflection gain reaches its maximum and thus, the cascaded channel can be written as
\begin{align}\label{eq:h_far}
	h(r_1,r_2)^{(\mathrm{far})} =&  \frac{l \cos\theta}{4\pi r_1 r_2} e^{jk(r_1+r_2)}. 
\end{align}

However, when the SRS width is excessively large or the transceivers are too close, the far-field assumption breaks down and the signals will not add perfectly in phase. Next, we analyze this near-field interference effect. Using the second-order Taylor expansion $\sqrt{1+x}=1+\frac{x}{2}-\frac{x^{2}}{8}+\mathcal{O}(x^{3})$, $d_1(x)$ and $d_2(x)$ can be approximated as
\begin{align}
	d_1(x)&=\sqrt{(r_1\sin\theta+x)^2+(r_1\cos\theta)^2}=r_1\sqrt{1+2\sin\theta\frac{x}{r_1}+\left(\frac{x}{r_1}\right)^2}\nonumber\\&\approx r_1 + x\sin\theta + \frac{\cos^2\theta}{2r_1}\,x^2, \\
	d_2(x)&=\sqrt{(r_2\sin\theta-x)^2+(r_2\cos\theta)^2}=r_2\sqrt{1-2\sin\theta\frac{x}{r_2}+\left(\frac{x}{r_2}\right)^2}\nonumber\\&\approx r_2 - x\sin\theta + \frac{\cos^2\theta}{2r_2}\,x^2,
\end{align}
for $|x|\ll r_1,r_2$. Then, the phase $\phi(x)$ can be expressed as
\begin{equation}
\phi(x)=	d_1(x)+d_2(x)= r_1+r_2+\frac{\cos^2\theta}{2}\!\left(\frac{1}{r_1}+\frac{1}{r_2}\right)x^2,
	\label{eq:phase_quad}
\end{equation}
while the amplitude is approximated by \(A(x)\approx A(0)=\frac{\cos\theta}{4\pi r_1 r_2}\).
Substituting Eq.~(\ref{eq:phase_quad}) into Eq.~(\ref{eq:h_def}), the cascaded channel can be rewritten as
\begin{align}
		&h(r_1,r_2,\theta) = \frac{e^{jk(r_1+r_2)}\cos\theta}{4\pi r_1 r_2}
	\int_{-l/2}^{\,l/2}\!
	\exp\!\left\lbrace \frac{j k\cos^2\theta}{2}\!\left(\frac{1}{r_1}+\frac{1}{r_2}\right) x^2\right\rbrace \mathrm{d}x, \nonumber \\
		&\overset{\text{(a)}}= \frac{e^{jk(r_1+r_2)}\cos\theta}{4\pi r_1 r_2}\sqrt{\frac{4\pi}{k\cos^2\theta\left(\frac{1}{r_1}+\frac{1}{r_2}\right)}}\int_{0}^{\sqrt{\tfrac{k\cos^2\theta}{4\pi}\left(\frac{1}{r_1}+\frac{1}{r_2}\right)}l} e^{j \frac{\pi}{2} t^2} \mathrm{d}t \nonumber 
\end{align}
\begin{align}\label{eq:h_Fresnel_form}
	&\overset{\text{(b)}}= \frac{e^{jk(r_1+r_2)}\cos\theta}{4\pi r_1 r_2}\;
	\frac{1}{\beta}\;
	\Bigg[
	C\!\left(\beta l\right)
	+ j\,S\!\left(\beta l\right)
	\Bigg]
\end{align}
where $\beta :
=\sqrt{\frac{k\cos^2\theta}{4\pi}\Big(\frac{1}{r_1}+\frac{1}{r_2}\Big)}$, (a) is realized by letting $t=\sqrt{\tfrac{k\cos^2\theta}{\pi}\left(\frac{1}{r_1}+\frac{1}{r_2}\right)}\,x$ and (b) is realized by using Fresnel integrals 
\(
C(v):=\int_0^{v}\cos\!\left(\tfrac{\pi}{2}u^2\right)\mathrm{d}u,\;
S(v):=\int_0^{v}\sin\!\left(\tfrac{\pi}{2}u^2\right)\mathrm{d}u
\)	\cite{1137900}. Then, the cascaded channel gain can be obtained as 
\begin{align}\label{eq:g_Fresnel_form}
	g(r_1,r_2,\theta)
	&= \big|h(r_1,r_2,\theta)\big|^2 = \frac{\cos^2\theta}{16\pi^2 r_1^2 r_2^2}\,\frac{1}{\beta^2}\,
	\Big(C(\beta l)^2 + S(\beta l)^2\Big).
\end{align}

\subsection{Asymptotic Analysis}
Using the asymptotic limits of the Fresnel functions, \(C(v)\approx v\) and \(S(v)\approx 0\) as \(v\to 0\), we obtain the same far-field channel expression as in Eq.~(\ref{eq:h_far}), along with the corresponding channel-gain expression:
\begin{align}\label{eq:g_far}
	g^{(\mathrm{far})}(r_1,r_2)
	= \big|h^{(\mathrm{far})}(r_1,r_2)\big|^2
	= \frac{l^{2}\cos^2\theta}{16\pi^{2} r_1^{2} r_2^{2}}.
\end{align}

Another interesting case is the limit where the SRS width tends to infinity, which corresponds to continuously deploying direction-aligned SRS panels along both sides of a roadway to realize continuous-area signal enhancement; for a user located on the roadway, the SRS can be regarded as infinitely wide. 

Noting that \(x=0\) is the unique stationary point of the phase \(\phi(x)\), the value of the improper integral is dominated by the adjacent domain of this stationary point. As \(l\to\infty\), the integral can therefore be approximated by the contribution around \(x=0\), and we can analyze it by directly taking the limit of Eq.~\eqref{eq:h_Fresnel_form}.
The channel and gain can then be computed as
\begin{align}
	h^{(\infty)}(r_1,r_2,\theta)
	&= \lim_{l\to\infty}\frac{e^{jk(r_1+r_2)}\cos\theta}{4\pi r_1 r_2}\,\frac{1}{\beta}\Big[C(\beta l)+j\,S(\beta l)\Big]\nonumber\\&
	\overset{\text{(a)}}= \frac{e^{jk(r_1+r_2)}\cos\theta}{4\pi r_1 r_2}\,\frac{1+j}{2\,\beta},\\
	g^{(\infty)}(r_1,r_2,\theta)
	&= \big|h^{(\infty)}(r_1,r_2,\theta)\big|^{2}
	= \frac{\cos^2\theta}{32\,\pi^{2} r_1^{2} r_2^{2}}\,\frac{1}{\beta^{2}},\label{eq:g_inifine_l} 
\end{align}
where (a) is realized by using the Fresnel limits \(C(v)\to\tfrac{1}{2}\), \(S(v)\to\tfrac{1}{2}\) as \(v\to\infty\) in Eq.~(7.3.20) of  \cite{mabramowitz64:handbook}.
\subsection{Maximum Value Analysis on $ g(r_1,r_2,\theta)$}
\subsubsection{With respect to width $l$}
An interesting question is at which value of $l$ the channel gain $g(r_1,r_2,\theta)$ reaches its maximum, i.e., 
\begin{equation}
	l^{*} = \arg\max_{l} \, g(r_1,r_2,\theta;l).
\end{equation}
For simplicity, we define $\mathcal{Q}(v) := C(v)^2 + S(v)^2$, where $v := \beta l$. Then, $g(r_1,r_2,\theta;l)
= \frac{1}{16\pi^2 r_1^2 r_2^2}\,\frac{1}{\beta^2}\,\mathcal{Q}(v)$ and the problem can be transformed as
\begin{equation}
	\max_{l\ge0} g(r_1,r_2,\theta;l) \quad \Longleftrightarrow \quad \max_{v\ge0}\mathcal{Q}(v).
\end{equation}
According to the first-order optimality conditions for $v$, the stationary points $v^*$ satisfies
\begin{equation}\label{eq:crit_v}
	C(v^*)\cos\!\left(\tfrac{\pi}{2}v^{* 2}\right)
	+ S(v^*)\sin\!\left(\tfrac{\pi}{2}v^{* 2}\right)=0,
\end{equation}
while the second derivative
\begin{equation}
	\mathcal{Q}''(v)
	= 2\Big[1+\pi v\big(-C(v)\sin(\tfrac{\pi}{2}v^2)
	+ S(v)\cos(\tfrac{\pi}{2}v^2)\big)\Big],
\end{equation}
is used to determine whether a stationary point is a maximum or minimum. Since Eq.~(\ref{eq:crit_v}) is a transcendental equation involving Fresnel integrals and trigonometric
functions of the same quadratic argument, no closed-form solution is known in terms of elementary or standard special-function inverses. Therefore, we provide a numerical solution as an approximation. Noting that $\mathcal{Q}(v)\to\tfrac{1}{2}$ as $v\to\infty$ and decays with oscillations, $\mathcal{Q}(v)$ exhibits a sequence of diminishing local maxima. Thus, the first radial maximum corresponds to the global maximum. Based on numerical evaluation of the radial first maximum of Eq.~\eqref{eq:crit_v}, we obtain
\begin{equation}
	v^* \approx 1.209378, \qquad
	\mathcal{Q}(v^*) \approx 0.90070817,
\end{equation}
and $\mathcal{Q}''(v^*)<0$, confirming that it is a maximum. Then the optimal width can be obtained as 
\begin{equation}\label{eq:maxgain_l}
	l^*= \frac{v^*}{\beta}
	\approx 1.209378\sqrt{\frac{4\pi}{k\cos^2\theta\left(\frac{1}{r_1}+\frac{1}{r_2}\right)}}.
\end{equation}
\subsubsection{With respect to angle $\theta$} 
The channel gain $g(r_1,r_2,\theta)$ can be rewritten as 
\begin{equation}
g(r_1,r_2,\theta)
=\frac{1}{16\pi^{2}r_1^{2}r_2^{2}}\,
\frac{4\pi}{k\left(\tfrac{1}{r_1}+\tfrac{1}{r_2}\right)}\;
\mathcal{Q}\left(\sqrt{\frac{k\cos^2\theta}{4\pi}\Big(\frac{1}{r_1}+\frac{1}{r_2}\Big)} l\right).
\end{equation}
Since $\mathcal{Q}(v)$ attains its global maximum at the first peak
$v^{*}\approx 1.209378$, the maximum angle $\theta^*$ can be obtained as:
\begin{equation}\label{eq:maxgain_theta}
\theta^*=	\begin{cases}
		\qquad \qquad 0, & l\sqrt{\dfrac{k}{4\pi}\!\left(\dfrac{1}{r_1}+\dfrac{1}{r_2}\right)}\le v^{*},\\[0.8em]
		\arccos\!\left(\dfrac{v^{*}}{\,l\sqrt{\dfrac{k}{4\pi}\left(\dfrac{1}{r_1}+\dfrac{1}{r_2}\right)}\,}\right), & \text{otherwise}.
	\end{cases}
\end{equation}
Therefore, when $r_1$ and $r_2$ are sufficiently large compared with the aperture width $l$, the maximum gain is achieved at the front side, i.e., at the direction $\theta=0$ to the surface. Otherwise, the maximum occurs at the other certain angle.

\subsection{Near-field and Far-field Boundary}
Since the classical phase-discrepancy-based near-field range $\frac{2l^{2}}{\lambda}$ relies on the paraxial approximation $\sin \theta \approx \theta$, $\cos\theta \approx 1   $ and is not applicable to reflecting scenarios, we study an SRS near–far-field boundary defined in terms of channel gain. This boundary captures the effective validity range of the far-field channel in Eq.~(\ref{eq:g_far}). By setting a small threshold $\delta$, the far-field range is determined by the following inequality:
\begin{equation}\label{eq:def_boundary}
	\frac{\big|g(r_1,r_2,\theta)-g^{(\mathrm{far})}(r_1,r_2)\big|}{g^{(\mathrm{far})}(r_1,r_2)}\ \le\ \delta
\end{equation} 
Substituting Eq.~(\ref{eq:g_Fresnel_form}) and Eq.~(\ref{eq:g_far}) into Eq.~(\ref{eq:def_boundary}) yields
\begin{equation}
	\left|\frac{\mathcal{Q}(v)}{v^2}-1\right|\ \le\ \delta.
\end{equation}
By using the power-series expansions of the Fresnel integrals in Eq.~(7.3.11) and Eq.~(7.3.13) of \cite{mabramowitz64:handbook}, for small $v$ we have
\begin{equation}
	\mathcal{Q}(v)=C(v)^2+S(v)^2
	= v^{2}-\frac{\pi^{2}}{45}\,v^{6}+\mathcal{O}(v^{10})
\end{equation}
Since $v=\beta l$ and $\beta^2=\dfrac{k\cos^2\theta}{4\pi}\!\left(\dfrac{1}{r_1}+\dfrac{1}{r_2}\right)$, we have 
\begin{equation}\label{eq:boundary_general}
		\frac{1}{r_1}+\frac{1}{r_2}\ \le\ \frac{12\sqrt{5\,\delta}}{k\,\cos^2\theta\,l^2}  ,
\end{equation}
which gives the far-field range in terms of $(r_1,r_2)$ for a given $(l,\theta,\lambda)$. Without loss of generality, setting $\delta=0.05$ in Eq.~\eqref{eq:boundary_general} gives a near–far-field boundary as
\begin{equation}\label{eq:boundary_delta005_k}
		\frac{1}{r_1}+\frac{1}{r_2}\ =\ \frac{6}{\,k\,\cos^{2}\theta\,l^{2}\,}\,.
\end{equation}
 In this regime,
the gain loss due to phase discrepancies across the surface can be regarded as negligible.

\section{Analysis of SRS: the Angle-misaligned Case}
Next, we investigate the case where the transmitter azimuth $\theta_{1}$ differs from the receiver azimuth $\theta_{2}$. This setting characterizes the co-channel interference observed by users at other locations, as well as the mainlobe width and sidelobe levels of the reflected beam. 
\subsection{Cascaded Channel and Channel Gain}
Assuming an ideal specular reflection, the cascaded channel between the transmitter and receiver via an SRS reflector can be expressed as
\begin{align}
	h(r_1,r_2,\theta_1,\theta_2)
	&=\!\! \int_{-l/2}^{\,l/2} A(x;\theta_1,\theta_2)\,
	\exp\!\left\{ j k \big[d_1(x;\theta_1)+ d_2(x;\theta_2)\big]\right\}\!\mathrm{d}x,
\end{align}
where $A(x;\theta_1,\theta_2):= \frac{\sqrt{r_1 \cos\theta_1 r_2 \cos\theta_2}}{4\pi d_2^{3/2}(x;\theta_2)d_1^{3/2}(x;\theta_1)}$, $d_1(x;\theta_1) := \sqrt{(r_1\sin\theta_{1}+x)^2+(r_1\cos\theta_{1})^2}$ and $d_2(x;\theta_2) := \sqrt{(r_2\sin\theta_{2}-x)^2+(r_2\cos\theta_{2})^2}$. Similarly, since \(|x| \ll r_1, r_2\), the amplitude factor \(A(x;\theta_1,\theta_2)\) can be approximated by its value \(A(0)= \frac{\sqrt{\cos\theta_1 \cos\theta_2}}{4\pi r_1 r_2}\) at the center. By using the second-order Taylor expansion, we have 
\begin{align}
	d_1(x;\theta_1)&=\sqrt{(r_1\sin\theta_1+x)^2+(r_1\cos\theta_1)^2}
	\nonumber \\& \approx\ r_1 + x\sin\theta_1 + \frac{\cos^2\theta_1}{2r_1}\,x^2,
\end{align}
\begin{align}
	d_2(x;\theta_2)&=\sqrt{(r_2\sin\theta_2-x)^2+(r_2\cos\theta_2)^2}
	\nonumber \\& \approx\ r_2 - x\sin\theta_2 + \frac{\cos^2\theta_2}{2r_2}\,x^2,
\end{align}
Then, the cascaded channel $h(r_1,r_2,\theta_1,\theta_2)$ can be rewritten as
\begin{align}\label{eq:h_mis_closed_form}
h(r_1,r_2,\theta_1,\theta_2)\!
&= \!\frac{e^{jk(r_1+r_2)}\sqrt{\cos\theta_1 \cos\theta_2}}{4\pi r_1 r_2}
\int_{-l/2}^{\,l/2}\!
e^{j\left(a\,x^{2}+b\,x\right)}\mathrm{d}x, \nonumber \\
\overset{\text{(a)}}=& \frac{e^{jk(r_1+r_2)}\sqrt{\cos\theta_1 \cos\theta_2}}{4\pi r_1 r_2}e^{-j\frac{b^{2}}{4a}}
\sqrt{\frac{\pi}{2a}}\int_{t_{-}}^{t_{+}} e^{j \frac{\pi}{2} t^2} \mathrm{d}t \nonumber \\
\overset{\text{(b)}}=&\frac{e^{jk(r_1+r_2)}\sqrt{\cos\theta_1 \cos\theta_2}}{4\pi r_1 r_2}\;
e^{-j\frac{b^{2}}{4a}}
\sqrt{\frac{\pi}{2a}}
\nonumber\\ &\times\Big(\,[C(t_{+})-C(t_{-})] + j\,[S(t_{+})-S(t_{-})]\,\Big),
\end{align}
where $a :=  \frac{k}{2}\left(\frac{\cos^{2}\theta_{1}}{r_{1}}+\frac{\cos^{2}\theta_{2}}{r_{2}}\right)$, $ b :=k\big(\sin\theta_{1}-\sin\theta_{2}\big)$, (a) is achieved by letting $t = \sqrt{\frac{2a}{\pi}}\left(x+\frac{b}{2a}\right)$ and $t_{\pm}= \sqrt{\frac{2a}{\pi}}\left(\pm\frac{l}{2}+\frac{b}{2a}\right)$, (b) is obtained by using Fresnel integrals. Then, the channel gain can be expressed as
\begin{align}\label{eq:g_mis_closed_form}
&g(r_1,r_2,\theta_1,\theta_2)=	|h(r_1,r_2,\theta_1,\theta_2)|^{2} \nonumber\\&\!\!=\!\! \frac{{\cos\theta_1 \cos\theta_2}}{16\pi^{2} r_1^{2} r_2^{2}}\;\frac{\pi}{2a}\;
	\Big\{\,[C(t_{+})-C(t_{-})]^{2}+[S(t_{+})-S(t_{-})]^{2}\,\Big\}.
\end{align}

\subsection{Asymptotic Analysis}
Under the far-field assumption $a\,l^{2}\ll 1$ and neglecting the quadratic term, the channel and the corresponding gain can be obtained as:
\begin{align}
	h^{(\mathrm{far})}(r_1,r_2,\theta_{1},\theta_{2})&= \frac{e^{jk(r_1+r_2)}\sqrt{\cos\theta_1 \cos\theta_2}}{4\pi r_1 r_2}\int_{-l/2}^{l/2}e^{j b x}\mathrm{d}x
	\nonumber\\&=\frac{e^{jk(r_1+r_2)}\sqrt{\cos\theta_1 \cos\theta_2}}{2b\pi r_1 r_2}\sin(\tfrac{b l}{2})
\end{align}
\begin{equation}\label{eq:g_mis_far}
	g^{(\mathrm{far})}(r_1,r_2,\theta_{1},\theta_{2})
	= \frac{{\cos\theta_1 \cos\theta_2}}{4k(\sin\theta_{1}-\sin\theta_{2})^2\pi^{2} r_1^{2} r_2^{2}}\sin^2(\tfrac{k(\sin\theta_{1}-\sin\theta_{2}) l}{2})
\end{equation}
which implies that the interfering signal will not experience a focusing gain in the far field. Instead, it fluctuates with a small amplitude in a sinc‑type manner.

For the case when \(l\to\infty\), since \(x=0\) is no longer the stationary point of the phase \(\phi(x;\theta_1,\theta_2)\), we cannot analyze the limit directly from Eq.~\eqref{eq:h_mis_closed_form}. When \(\theta_{1}\neq\theta_{2}\), the stationary point $x_{\mathrm{sp}}$ is obtained from the first‑order condition of \(\phi(x;\theta_1,\theta_2)=d_1(x;\theta_1)+d_2(x;\theta_2)=\sqrt{(r_1\sin\theta_1+x)^2+(r_1\cos\theta_1)^2}+\sqrt{(r_2\sin\theta_2-x)^2+(r_2\cos\theta_2)^2}\). Then, we have
\begin{align}
	\phi'(x_{\mathrm{sp}};\theta_1,\theta_2)
	&= \frac{x_{\mathrm{sp}}+r_1\sin\theta_1}{d_1(x_{\mathrm{sp}};\theta_1)}+\frac{x_{\mathrm{sp}}-r_2\sin\theta_2}{d_2(x_{\mathrm{sp}};\theta_2)}=0.
	\label{eq:SP_eq}
\end{align}
Since
\begin{equation}
	\phi''(x;\theta_1,\theta_2)=\frac{(r_1\cos\theta_1)^2}{d_1(x;\theta_1)^3}+\frac{(r_2\cos\theta_2)^2}{d_2(x;\theta_2)^3}>0,
\end{equation}
 $\phi'(x;\theta_1,\theta_2)$ is strictly increasing on $\mathbb{R}$, hence the unique $x_{\mathrm{sp}}$ can be obtained by solving Eq.~(\ref{eq:SP_eq}) as
\begin{equation}\label{eq:xsp-closed}
		x_{\mathrm{sp}}=\frac{r_1 r_2\,\sin(\theta_2-\theta_1)}{\,r_1\cos\theta_1+r_2\cos\theta_2\,}.
\end{equation}
For simplicity, we denote $d_{1,\mathrm{sp}}:= d_1(x_{\mathrm{sp}};\theta_1)$, $
d_{2,\mathrm{sp}}:= d_2(x_{\mathrm{sp}};\theta_2)$, $
\phi_{\mathrm{sp}}:= \phi(x_{\mathrm{sp}};\theta_1,\theta_2)$ and $
\phi''_{\mathrm{sp}}:= \phi''(x_{\mathrm{sp}};\theta_1,\theta_2)$. Then, the
Taylor expansion for the phase at $x_{\mathrm{sp}}$ is
\begin{equation}
	\phi(x;\theta_1,\theta_2)=\phi_{\mathrm{sp}}+\frac{1}{2}\phi''_{\mathrm{sp}}(x-x_{\mathrm{sp}})^2
	+\mathcal{O}\big((x-x_{\mathrm{sp}})^3\big),
\end{equation}
and the amplitude $A(x)$ can be approximated by $ A(x_{\mathrm{sp}})=\frac{\sqrt{r_1\cos\theta_1\,r_2\cos\theta_2}}{4\pi\,d_{1,\mathrm{sp}}^{3/2}d_{2,\mathrm{sp}}^{3/2}}$. Then, the channel can be obtained as
\begin{align}
	h(r_1,r_2,\theta_1,\theta_2)^{\left( \infty\right) }&= A(x_{\mathrm{sp}})\,e^{\,jk\phi_{\mathrm{sp}}}
	\int_{-\infty}^{\infty}\exp\!\left\{j k\frac{\phi''_{\mathrm{sp}}}{2}
	\,(x-x_{\mathrm{sp}})^2\right\}\mathrm{d}x \nonumber \\
	&\overset{\text{(a)}}=A(x_{\mathrm{sp}})\,e^{\,jk\phi_{\mathrm{sp}}}\,
	e^{\,j\pi/4}\,\sqrt{\frac{2\pi}{k\,\phi''_{\mathrm{sp}}}}.
\end{align}
where $(a)$ is obtained by using $ \int_{0}^{\infty}e^{-t^2}d\mathrm{t}=\sqrt{\pi}/2$ in Eq.~(7.4.1) of  \cite{mabramowitz64:handbook}. Then, the cascaded channel gain can be expressed as
\begin{align}\label{eq:g_inf_generic}
		g^{\left( \infty\right) }(r_1,r_2,\theta_1,\theta_2)
		&=\frac{2\pi}{k\,\phi''_{\mathrm{sp}}}\;\big|A(x_{\mathrm{sp}})\big|^{2} \nonumber\\&=\frac{r_1 r_2\,\cos\theta_1\cos\theta_2}
		{8\pi\,k\,
			\Big((r_1\cos\theta_1)^2d_{2,\mathrm{sp}}^{3}
			+(r_2\cos\theta_2)^2 d_{1,\mathrm{sp}}^{3}\Big)\,}.
\end{align}
Furthermore, using the quadratic expansion of the phase at $x_{\mathrm{sp}}$ and taking $A(x)\approx A(x_{\mathrm{sp}})$ in the stationary region, we have
\begin{align}\label{eq:h_sp_finite}
	&h^{(\mathrm{sp})}(r_1,r_2,\theta_1,\theta_2)=
	A(x_{\mathrm{sp}})\,e^{jk\phi_{\mathrm{sp}}}
	\int_{-l/2}^{\,l/2}
	\exp\!\left\{j\,\frac{k\,\phi''_{\mathrm{sp}}}{2}\,(x-x_{\mathrm{sp}})^2\right\}\mathrm{d}x.\nonumber \\
	&=A(x_{\mathrm{sp}})\,e^{jk\phi_{\mathrm{sp}}}\;
	\sqrt{\frac{\pi}{k\,\phi''_{\mathrm{sp}}}}\;
	\Big(\,[C(t^{(\mathrm{sp})}_{+})-C(t^{(\mathrm{sp})}_{-})]+j\,[S(t^{(\mathrm{sp})}_{+})-S(t^{(\mathrm{sp})}_{-})]\,\Big).
\end{align}
where $t^{(\mathrm{sp})}_{\pm}=\sqrt{\frac{k\,\phi''_{\mathrm{sp}}}{\pi}}\!\left(\pm\frac{l}{2}-x_{\mathrm{sp}}\right)$. Then, $g^{(\mathrm{sp})}(r_1,r_2,\theta_1,\theta_2)$ can be obtained as 
\begin{align}\label{eq:g_sp_finite}
	&g^{(\mathrm{sp})}(r_1,r_2,\theta_1,\theta_2)\nonumber \\ &\approx
	\big|A(x_{\mathrm{sp}})\big|^{2}\;\frac{\pi}{k\,\phi''_{\mathrm{sp}}}\;
	\Big\{[C(t^{(\mathrm{sp})}_{+})-C(t^{(\mathrm{sp})}_{-})]^{2}+[S(t^{(\mathrm{sp})}_{+})-S(t^{(\mathrm{sp})}_{-})]^{2}\Big\}.
\end{align}
The approximation Eq.~\eqref{eq:h_sp_finite}–Eq.~\eqref{eq:g_sp_finite} is most accurate when the stationary point is well contained inside the aperture, i.e., $x_{\mathrm{sp}}\in(-l/2,l/2)$ and not too close
to an edge.


\subsection{Beam Pattern and Beamwidth}
The conventional far-field beam pattern is obtained from the steering vector and does not need to be normalized by path loss, since there is no distance resolution. Near-field beam patterns are often normalized by path loss to illustrate the energy-focusing effect \cite{10721321,11212817}. For an SRS, the received signal strength is affected by path loss, coherent signal combining, and power projection. By normalizing the beam pattern with respect to path loss, we can reveal the effective operating region of the SRS in both the near and far field, where the signals add mostly constructively and the power projection is large. The SRS beam pattern function is defined via the path-loss–normalized channel gain as

{
	\setlength{\abovedisplayskip}{-8pt}
	\setlength{\abovedisplayshortskip}{0pt}
\begin{equation}\label{eq:41}
	F(r_2,\theta_2; r_1,\theta_1)\!=\!\cos\theta_1 \cos\theta_2\frac{\pi}{2a}
	\Big\{\,[C(t_{+})-C(t_{-})]^{2}\!+[S(t_{+})-S(t_{-})]^{2}\Big\}.
\end{equation}}
When $r_2$ is larger than a certain threshold, $F(r_2,\theta_2; r_1,\theta_1)$ attains its maximum at $\theta_2 = \theta_1$, i.e., the main beam is formed in the direction $\theta_2 = \theta_1$. Then, the mainlobe width can be defined according to the 3-dB beamwidth as
\begin{equation}\label{eq:3db}
	\mathrm{BW}^{(\mathrm{3dB})} = \theta_2^{(\mathrm{max})} -\theta_2^{(\mathrm{min})},
\end{equation}
where $ \theta_2^{(\mathrm{max})}$ and $\theta_2^{(\mathrm{min})}$ are the two solutions of
\begin{equation}\label{eq:theta_2}
		F(r_2,\theta_2; r_1,\theta_1)=\frac{1}{2}F(r_2,\theta_1; r_1,\theta_1).
\end{equation}

\subsubsection{Far-field closed-form 3-dB beamwidth}
Based on Eq.~(\ref{eq:g_mis_far}), the beam pattern in far-field can be written as
\begin{equation}\label{eq:f_mis_far}
	F(r_2,\theta_2; r_1,\theta_1)
	= 2 l\cos\theta_1\cos\theta_2
	\left[
	\frac{\sin\!\big(\tfrac{k l}{2}(\sin\theta_2-\sin\theta_1)\big)}
	{\tfrac{k l}{2}(\sin\theta_2-\sin\theta_1)}
	\right]^2.
\end{equation}
Substituting Eq.~(\ref{eq:f_mis_far}) into Eq.~(\ref{eq:theta_2}), we have
\begin{equation}\label{eq:45}
	2\cos \theta_2 \left(
	\frac{\sin u_{3\mathrm{dB}}}{u_{3\mathrm{dB}}}
	\right)^2= \cos \theta_1
\end{equation}
where $u_{3\mathrm{dB}}:=\frac{k l}{2}\big(\sin\theta_2-\sin\theta_1\big)$. Near the main beam ($\theta_2\approx\theta_1$), we have $\cos\theta_2\approx\cos\theta_1$. Then, $u_{3\mathrm{dB}}$ can be numerically obtained by Eq~(\ref{eq:45}) as  $u_{3\mathrm{dB}}\approx 1.3916 $ rad. Since both $\theta_2^{(\mathrm{max})}$ and $\theta_2^{(\mathrm{min})}$ are close to $\theta_1$, 
by a first–order Taylor expansion around $\theta_1$ we obtain
\begin{equation}\label{eq:46}
\sin\theta_2^{(\mathrm{max})} - \sin\theta_2^{(\mathrm{min})}
\approx \cos\theta_1\left(\theta_2^{(\mathrm{max})} - \theta_2^{(\mathrm{min})}\right).
\end{equation}
Using Eq.~(\ref{eq:46}) and $u_{3\mathrm{dB}}=\frac{k l}{2}\big|\sin\theta_2-\sin\theta_1\big|$, the 3-dB beamwidth in far-field can be obtained as
\begin{equation}
		\mathrm{BW}^{(\mathrm{3dB})}
		\approx
		\frac{0.886\,\lambda}{l\cos\theta_1}.	\label{eq:3dB_bw_far_field_approx_en}
\end{equation}

\subsubsection{ Near-field 3-dB beamwidth} 
Substituting Eq.~(\ref{eq:41}) into Eq.~(\ref{eq:theta_2}), the 3-dB angles $\theta_2^{(\mathrm{max})}$ and $\theta_2^{(\mathrm{min})}$  are the two real roots of 
\begin{align}\label{eq:3dB_nearfield_equation_en}
\Phi(\theta_2)&=[C(t_{+})-C(t_{-})]^{2}+[S(t_{+})-S(t_{-})]^{2} \nonumber\\&-  \frac{2 a\cos\theta_{1}}{a_0\cos\theta_{2}} \Big(C(t_{0})^2 + S(t_{0})^2\Big)=0,
\end{align}
where $a_0 :=  \frac{k \cos^{2}\theta_{1}}{2}\left(\frac{1}{r_{1}}+\frac{1}{r_{2}}\right)$ and $ t_0 :=l\sqrt{\frac{a_0}{2\pi}}$, with $\theta_2^{(\min)}<\theta_1<\theta_2^{(\max)}$. Then, the near-field 3-dB main-lobe width is obtained by Eq.~(\ref{eq:3db}). Since Eq.~(\ref{eq:3dB_nearfield_equation_en}) is a transcendental equation involving Fresnel integrals and trigonometric functions, the closed-form solution in terms of elementary functions is hard to obtain. However, it is one-dimensional, and when $r$ is sufficiently large, it has a unique value for $\theta_2^{(\min)}<\theta_1<\theta_2^{(\max)}$. Consequently, standard scalar root-finding algorithms, such as bisection and Newton methods can be employed to compute $\theta_2^{(\min)}$ and $\theta_2^{(\max)}$ with arbitrary numerical accuracy. 

\subsubsection{The effective region of 3-dB beamwidth}
When the receiver is very close to an SRS, the main lobe located at $\theta_2 = \theta_1$ is no longer the maximum and may even vanish. In the following, we analyze this threshold to reveal the effective usage range of the 3-dB beamwidth. Since $F(r_2,\theta_2; r_1,\theta_1)$ in Eq.~\eqref{eq:41} involves both the near-field geometric curvature factor and the Fresnel pattern, it is difficult to solve for the actual threshold $r_{2,\mathrm{th}}$ at which the main lobe ceases to be the maximum. As the beam pattern varies slowly with the near-field geometric curvature factor $\cos\theta_1 \cos\theta_2\frac{\pi}{2a} $, we first carry out the analysis based on the Fresnel pattern, which is given by
\begin{equation}\label{eq:fresnel_pattern}
	\bar{F}(r_2,\theta_2; r_1,\theta_1)=
[C(t_{+})-C(t_{-})]^{2}+[S(t_{+})-S(t_{-})]^{2}
\end{equation}
Since $\bar{F}'(r_2,\theta_2; r_1,\theta_1)|_{\theta_2 = \theta_1}=0$, whether the main lobe is a local maximum is determined by the value of the second derivative $\bar{F}''(r_2,\theta_2; r_1,\theta_1)$. The range of $r_2$ for which a local maximum is attained at $\theta_2 = \theta_1$ can be obtained by solving the following equation:
\begin{equation}\label{eq:50}
	\bar{F}''(r_2,\theta_2; r_1,\theta_1)|_{\theta_2 = \theta_1}=8\pi u \left[ \,S(u)\cos\big(\tfrac{\pi}{2}u^{2}\big)-C(u)\sin\big(\tfrac{\pi}{2}u^{2}\big)\right]  < 0,
\end{equation}
where $u:= \sqrt{\frac{k}{4\pi}\left(\frac{\cos^{2}\theta_{1}}{r_{1}}+\frac{\cos^{2}\theta_{2}}{r_{2}}\right)}l$ and the smallest positive solution of $u$ in Eq.~\eqref{eq:50} can be numerically calculated as 1.6345. Then, the $r_{2,\mathrm{th}}$ can be obtained as
\begin{equation}\label{eq:51}
r_{2,\mathrm{th}} > \frac{1}{\frac{5.3437\mu\lambda}{l^{2}\cos^{2}\theta_1}-\frac{1}{r_1}} .
\end{equation}
where $\mu$ is an introduced factor related to the near-field geometric curvature factor. Numerically, choosing $\mu = 0.87$ provides a good approximation to the effective range.


\section{SRS aided Communication Systems}
In this section, we establish a single SRS-aided communication system to demonstrate the advantages of SRS. Typically, an SRS employs a motor-driven rotation to achieve beam alignment. Suppose that, at the initial position, the angles of incidence and departure of the transmitter and receiver with respect to the SRS surface normal are $\theta_1$ and $\theta_2$, and their distances are $r_1$ and $r_2$, respectively. The motor then rotates by an angle $\psi$ in the direction that increases the incidence angle. The transmitted symbol $ s $ follows a complex Gaussian distribution, \textit{i.e.}, $ s \sim \mathcal{CN}(0,1) $. Then, the received signal can be expressed as
\begin{align}\label{eq:received signal}
	y_k &=\sqrt{P_t} h(r_1,r_2,\theta_1-\psi,\theta_2+\psi)s+n_k, 
\end{align}
where $ P_t $ is the transmit power. The achievable rate can be expressed as
\begin{equation} \label{eq:rate}
	R=\log_2\left( 1+\frac{P_t}{\sigma^2}g(r_1,r_2,\theta_1-\psi,\theta_2+\psi)\right) .
\end{equation}
Without loss of generality, we assume that the locations of the transmitter and receiver satisfy Eq.~(\ref{eq:51}), i.e., the reflected beam of the SRS attains its maximum gain at the main-lobe direction $\theta_2 = \theta_1$. Then, the optimal $\psi^*$ can be obtained as
\begin{equation} \label{eq:optimal_psi}
	\psi^*=\max_{\psi} R = \frac{1}{2}(\theta_2-\theta_1),
\end{equation}
and the optimal achievable rate is 
\begin{equation} \label{eq:optimal_rate}
	R^*=\log_2\left( 1+\frac{P_t}{\sigma^2}g(r_1,r_2,\frac{\theta_1+\theta_2}{2})\right) .
\end{equation}
where $g(r_1,r_2,\frac{\theta_1+\theta_2}{2})$ can be obtained by Eq.~(\ref{eq:g_Fresnel_form}). Moreover, to better serve nearby users, the SRS can be partitioned into smaller SRS panels and assembled into an SRS array. 

Furthermore, since $\mathrm{projected\ power}\propto \cos(\theta_1-\psi)\cos(\theta_2+\psi)$, the rotation angle that maximizes the projected power is $\psi^{\star} = \arg\max_{\psi}\big[\cos(\theta_1-\psi)\cos(\theta_2+\psi)\big] = \frac{\theta_1+\theta_2}{2} + k\pi,\quad k \in \mathbb{Z}$, which exactly coincides with $\psi^{*}$. Therefore, by adopting the rotational control scheme, a larger projected power can be achieved compared to an ideal RIS.

\section{Numerical Results}
In this section, we conduct numerical performance analysis to demonstrate the channel gain and its asymptotic behavior of SRS under both aligned and misaligned configurations, as well as the near-/far-field boundary, beam pattern, and performance in wireless communication systems. As illustrated in Fig.~\ref{Single_SRS}, the center of the SRS is located at the origin, and the SRS is placed along the $x$-axis.
\subsection{Angle-aligned case}
We first analyze the performance of SRS under the aligned configuration. Unless otherwise specified, we set the wavelength $\lambda = 0.1\,\mathrm{m}$, the transceiver--SRS and SRS--receiver distances $r_1 = r_2 = 20\,\mathrm{m}$, and the incidence/reflection angle $\theta = 30^\circ$, and the width of SRS $l = 2\,\mathrm{m} $. The \emph{Calculated} curve is obtained by numerically evaluating $g=|h|^2$ by Eq.~(\ref{eq:h_def}), while the \emph{Estimated} curve uses the close-form expression in Eq.~(\ref{eq:h_Fresnel_form}) and Eq.~(\ref{eq:g_Fresnel_form}). We also include the \emph{Far-field} curve using Eq.~(\ref{eq:h_far}) for the small-$l$ and the asymptotic limit in Eq.~(\ref{eq:g_inifine_l}) for the large-$l$, respectively.

\vspace{-4pt}
\begin{figure}[htbp]
	\centering
	\includegraphics [width=70mm]{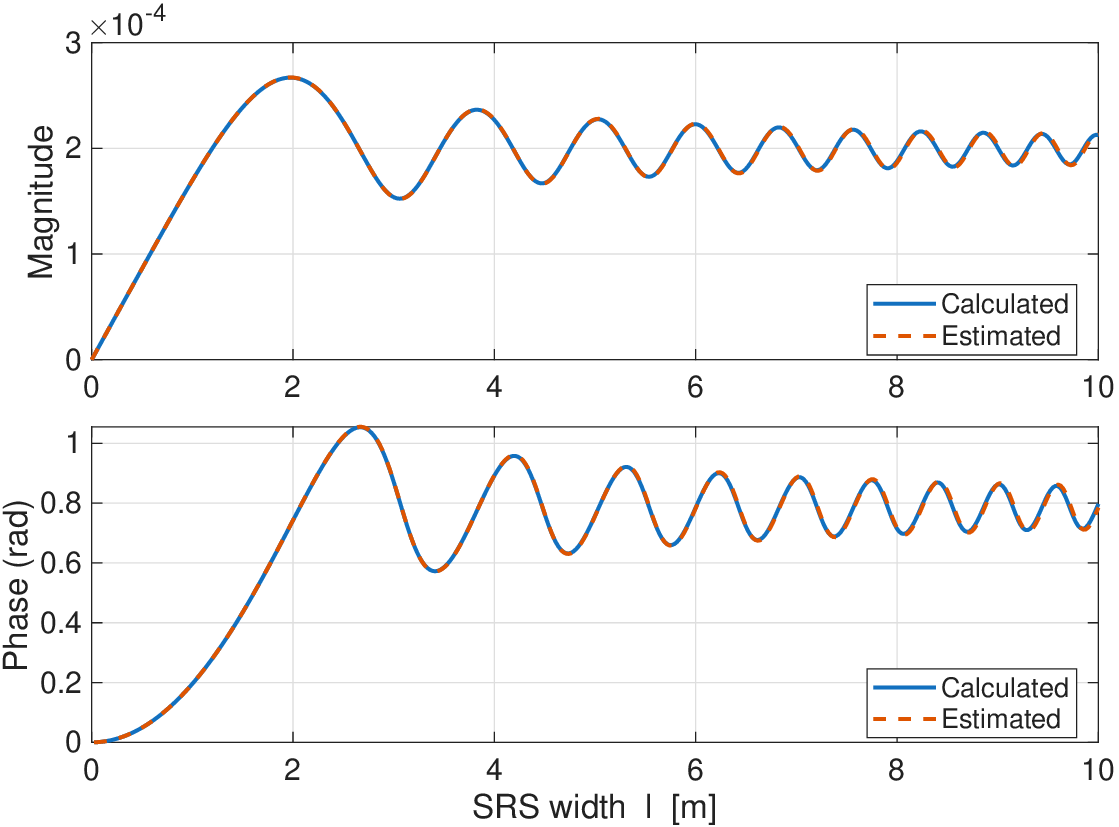}\\
	\caption{Magnitude and phase of channel vs width of SRS $l$.}
	\label{fig_srs_aligned_h_mag_phase_vs_l}
\end{figure}
We investigate the impact of the SRS width on the channel amplitude and on the phase deviation with respect to a reference phase $e^{-jk(r_1+r_2)}$ in Fig.~\ref{fig_srs_aligned_h_mag_phase_vs_l}. It can be observed from Fig.~\ref{fig_srs_aligned_h_mag_phase_vs_l} that the derived closed-form expressions for both the channel amplitude and phase accurately match the numerical results. As the SRS aperture increases, the channel amplitude and the relative phase first grows to their maximum values and then exhibit a damped oscillatory behavior, whose upper envelope gradually decreases and converges to a constant. This phenomenon can be explained as follows. When the transceiver is located in the far field of the SRS, all reflected components add up constructively. As the aperture further increases, the transceiver enters the near-field region, where the phases of the reflected signals from the edge region of the SRS vary periodically, leading to local reflection regions that are partially out of phase with the overall contribution. Meanwhile, the path loss associated with the edge reflections increases with the aperture size, which gradually reduces the oscillation magnitude.

\vspace{-4pt}
\begin{figure}[htbp]
	\centering  
	\subfigbottomskip=0pt 
	\subfigcapskip=-2pt 
	\subfigure[]{
		\includegraphics[width=70mm]{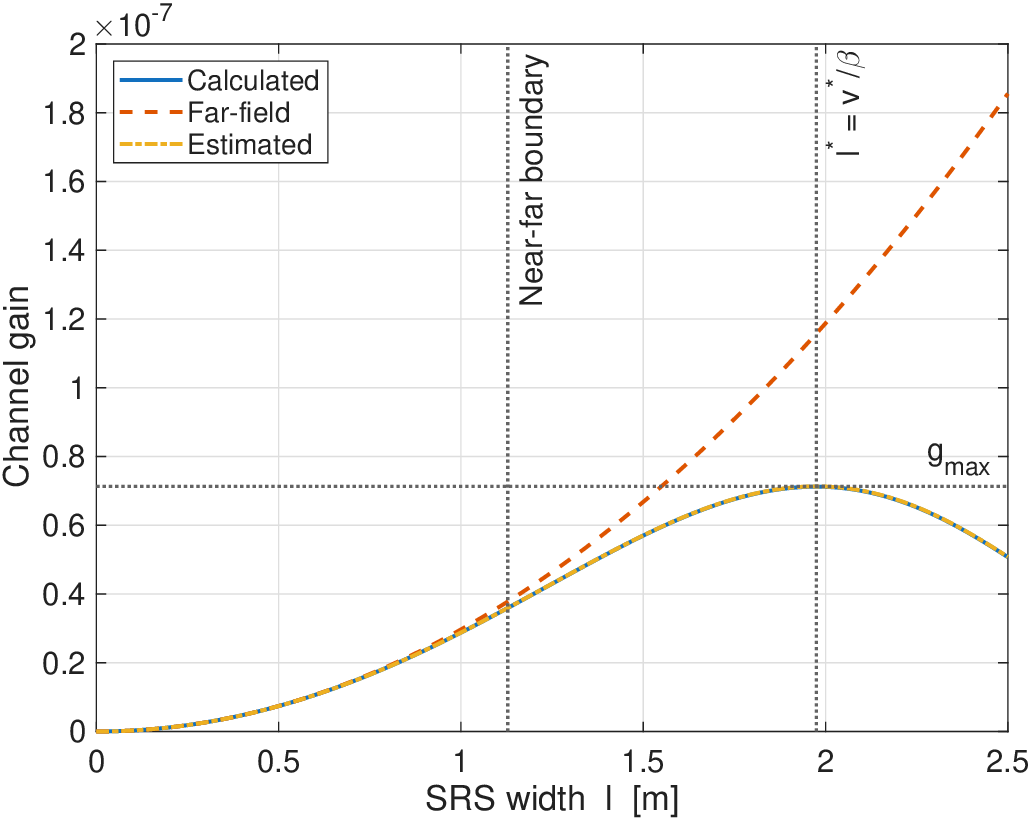} 
	}	\\	
	\subfigure[]{
		\includegraphics[width=70mm]{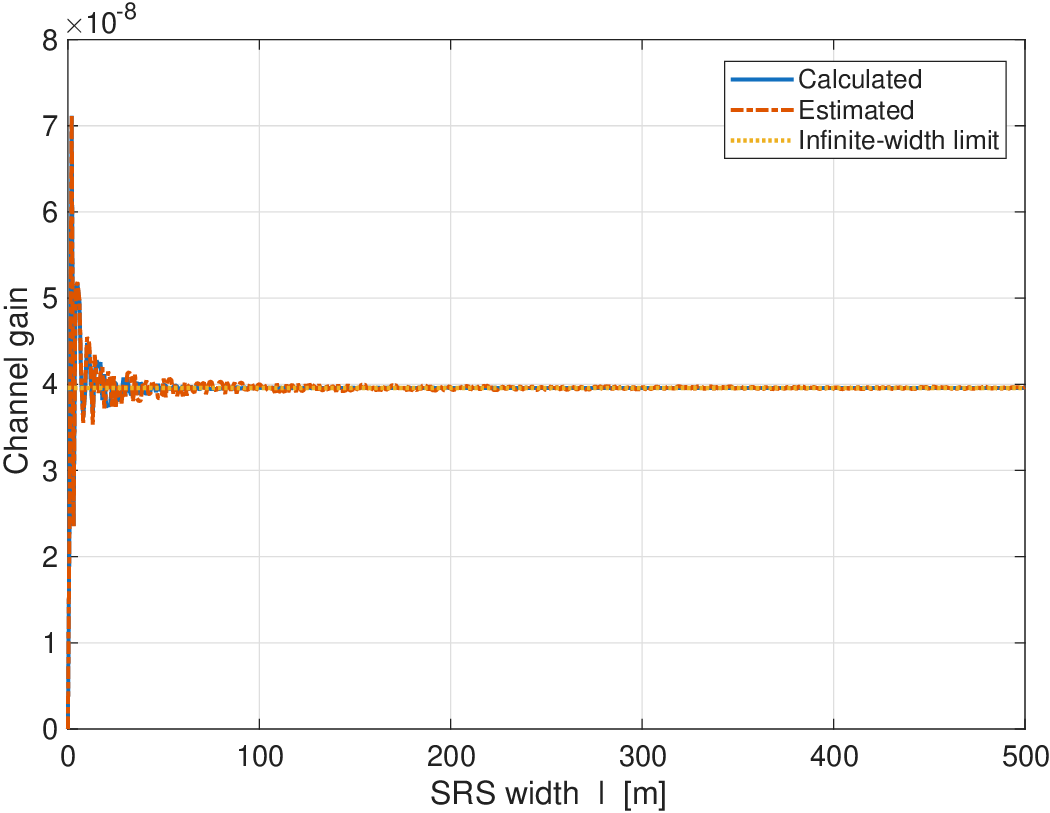} 
	}	
	\caption{Channel gain versus width of SRS $l$: (a) small aperture case, (b) large aperture case.}
	\label{fig_Channel_gain_vs_width}
	\vspace{-4pt}
\end{figure}
We characterize the impact of the SRS width on the channel gain in Fig.~\ref{fig_Channel_gain_vs_width}~(a). The near-field boundary given in Eq.~(\ref{eq:boundary_delta005_k}) and the aperture that achieves the maximum channel gain in Eq.~(\ref{eq:maxgain_l}) are indicated by dotted black lines. We observe from Fig.~\ref{fig_Channel_gain_vs_width}~(a) that the channel gain first increases and then decreases as the aperture size grows, which is due to the non-constructive combining of the reflected signals from the edge regions when the aperture becomes large. Moreover, the maximum gain is attained exactly at the aperture predicted by our derived expression for the optimal channel gain, thereby validating its correctness. In addition, the derived near-/far-field boundary accurately captures the transition between the two regimes. When the aperture is small and the transceiver lies in the far field, the channel gain coincides with the far-field approximation. As the aperture increases, the far-field gain gradually becomes inaccurate, since it fails to account for the non-constructive contributions from the edge reflections.

We investigate the asymptotic case where the SRS aperture tends to infinity in Fig.~\ref{fig_Channel_gain_vs_width}~(b). We notice from Fig.~\ref{fig_Channel_gain_vs_width}~(b) that, as the aperture increases, the accuracy of the Fresnel-based analytical expression gradually degrades. This is because only a second-order approximation is adopted; when the aperture becomes very large, higher-order analyses are required to maintain accuracy. In contrast, the derived closed-form expression in the asymptotic regime is shown to be accurate, revealing that the channel gain converges to a constant value that is independent of the aperture size. This behavior can be explained by the fact that, when the aperture is infinite, only the effective reflection region around the SRS center contributes significantly to the channel gain, while reflections from other regions have negligible impact due to increased path loss and position-dependent periodic phase variations. Physically, this implies that there always exists an effective reflection path, and the channel gain does not vanish as the aperture grows without bound.

\begin{figure}[htbp]
	\centering
	\includegraphics [width=70mm]{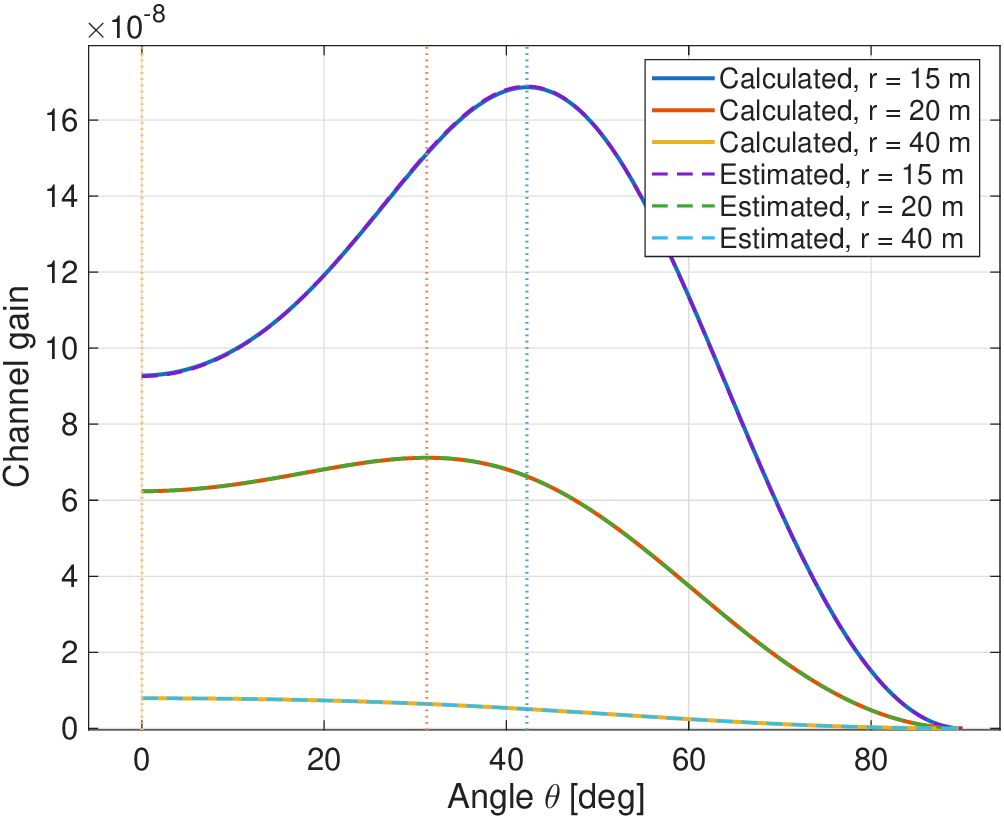}\\
	\caption{Channel gain vs angle $\theta$.}
	\label{single_SRS_gain_vs_angle}
\end{figure}
We characterize the impact of the incident/reflection angle on the channel gain in Fig.~\ref{single_SRS_gain_vs_angle}. The angle that achieves the maximum gain, as given in Eq.~(\ref{eq:maxgain_theta}), is indicated by dotted lines. We can observe from Fig.~\ref{single_SRS_gain_vs_angle} that, in the far-field regime, the channel gain is maximized when the incident and reflected rays are perpendicular to the SRS. As the system moves into the near-field regime, however, the optimal reflection angle gradually deviates from this perpendicular direction. This is because the effective aperture  gradually decreases as the angle increases. When the transceiver is sufficiently far from the SRS, reflections from all points on the surface combine approximately constructively, and increasing the angle only reduces the effective reflected power, so the maximum gain is achieved at normal incidence. When the transceiver is close to the SRS, a smaller effective aperture helps suppress non-constructive contributions from the edge regions, and thus the maximum gain is attained at a certain nonzero angle. Moreover, the derived analytical expression accurately captures the angle that maximizes the channel gain.

\begin{figure}[htbp]
	\centering
	\includegraphics [width=70mm]{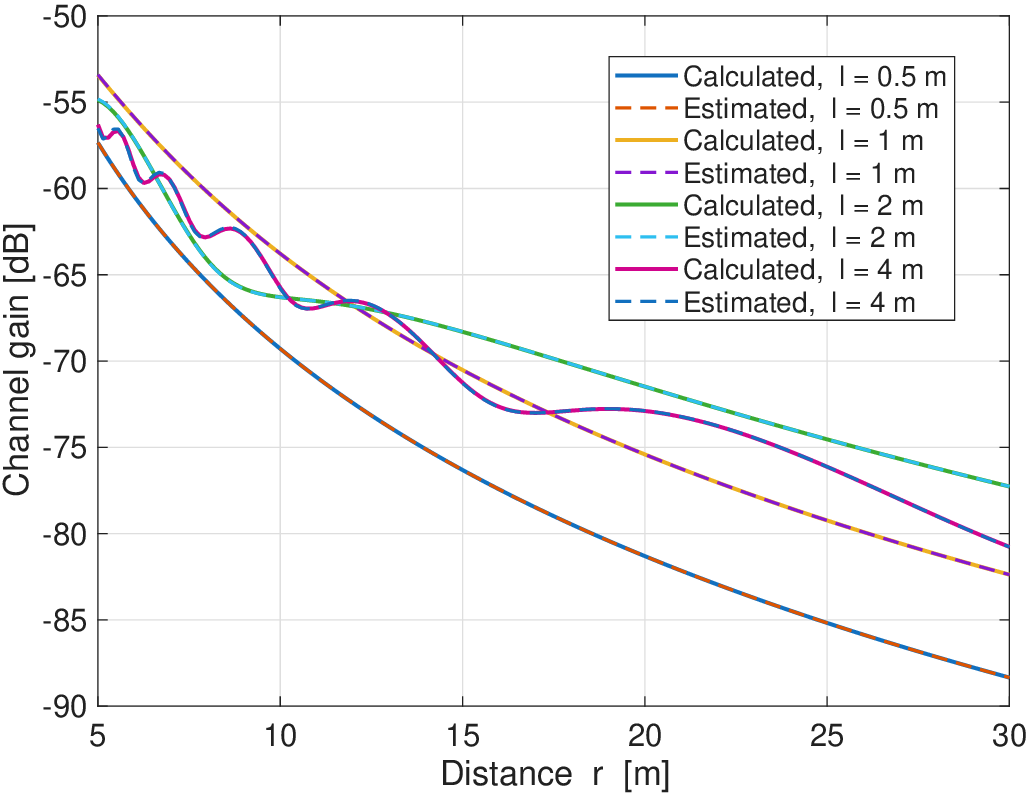}\\
	\caption{Channel gain vs distance $r$.}
	\label{single_SRS_gain_vs_distance}
\end{figure}
We investigate the impact of distance on the channel gain in Fig.~\ref{single_SRS_gain_vs_distance}. We observe from Fig.~\ref{single_SRS_gain_vs_distance} that the channel gain exhibits an overall decreasing trend as the distance increases. When the SRS aperture is small and the channel operates in the far-field regime, the gain decreases monotonically with distance. In contrast, when the SRS aperture is sufficiently large and the channel lies in the near-field regime, the channel gain decreases in a damped oscillatory manner. This is because the free-space propagation loss increases with distance, while in the near field the combination of reflected signals from different positions on the SRS varies periodically with the distance.

\begin{figure}[htbp]
	\centering
	\includegraphics [width=70mm]{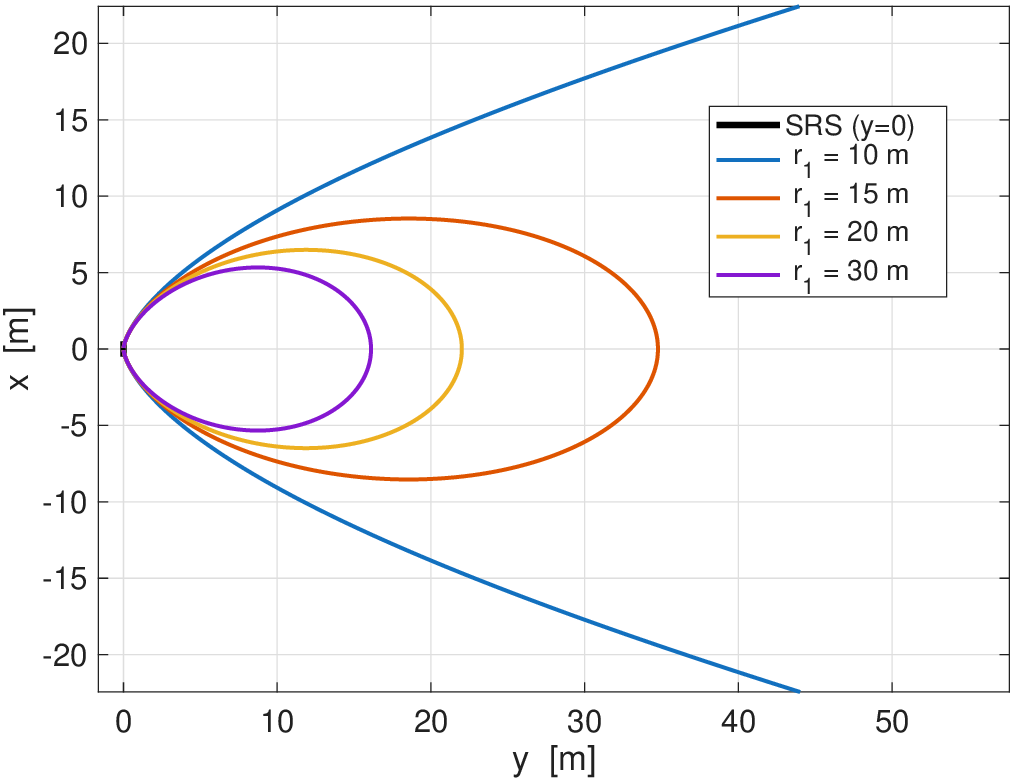}\\
	\caption{Near-far field boundary of the receiver.}
	\label{near-far_boundary}
\end{figure}
We characterize the near-/far-field boundary in Fig.~\ref{near-far_boundary}, where the SRS aperture is set to $1$~m. It can be observed from Fig.~\ref{near-far_boundary} that, when the transmitter is sufficiently far from the SRS, the near-/far-field boundary of the receiver exhibits a lemniscate-like shape; as the reflection angle increases, the boundary curve gradually moves closer to the SRS. When the transmitter is located closer to the SRS, however, the boundary curve becomes non-closed, and the near-field region extends to infinity for relatively small reflection angles. This is because the boundary is jointly determined by the transceiver distances to the SRS, the carrier frequency, and the SRS aperture. If either the transmitter or the receiver is very close to the SRS, the near-field non-constructive-combining effect becomes unavoidable in the cascaded channel.

\subsection{Angle-misaligned case}
Next, we analyze the performance of SRS under the angle-misaligned configuration. Unless otherwise specified, we set the incidence angle to $20^\circ$ and the reflection angle to $40^\circ$. The results obtained from the center-point-based closed-form expressions in Eq.~(\ref{eq:h_mis_closed_form}) and Eq.~(\ref{eq:g_mis_closed_form}) are labeled as \emph{CP-estimated}, while the curve obtained from the stationary-point-based closed-form expression in Eq.~(\ref{eq:g_sp_finite}) is labeled as \emph{SP-estimated}.

\begin{figure}[htbp]
	\centering
	\includegraphics [width=70mm]{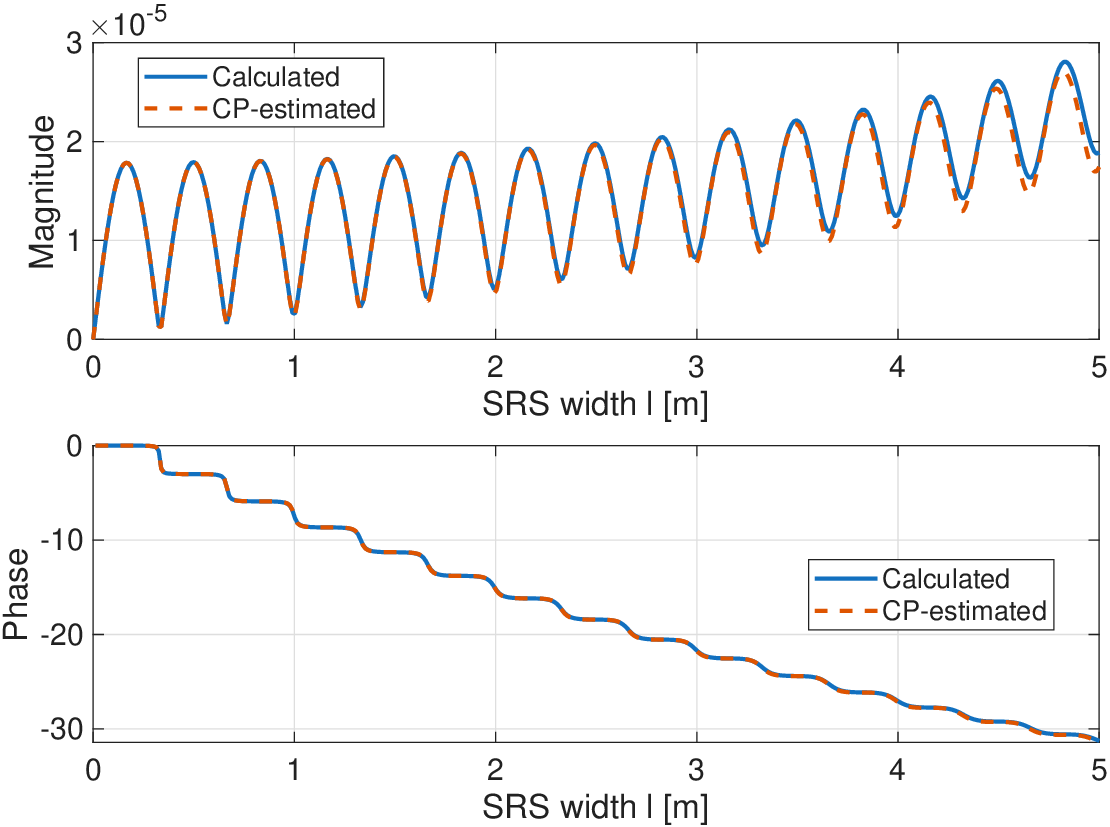}\\
	\caption{Channel vs width of SRS $l$.}
	\label{fig_srs_misaligned_h_mag_phase_vs_l}
\end{figure}
We investigate the impact of the SRS aperture on the angle-misaligned channel's amplitude and the phase deviation with respect to a reference phase $e^{-jk(r_1+r_2)}$ in Fig.~\ref{fig_srs_misaligned_h_mag_phase_vs_l}. We observe from Fig.~\ref{fig_srs_misaligned_h_mag_phase_vs_l} that the closed-form estimation based on the SRS midpoint is highly accurate for small apertures, whereas a noticeable deviation gradually appears for large apertures. This deviation stems from two reasons: i) as the aperture grows, higher-order approximations are required to maintain accuracy; and ii) unlike the aligned case, $x=0$ is no longer a stationary point of the phase function, and thus the approximation based on $A(0)$ becomes less accurate as the aperture increases. Moreover, as we increase the aperture, the channel amplitude starts from a relatively small value and then grows with oscillations. Meanwhile, the phase no longer oscillates back as in the aligned case; instead, it drifts away from the reference in a periodic manner. This is because the true stationary point of the phase now lies outside the physical aperture; as the aperture expands, it gradually approaches and eventually encompasses the stationary point, leading to periodically increasing oscillations in both the amplitude and the phase deviation.

\begin{figure}[htbp]
	\centering  
	\subfigbottomskip=0pt 
	\subfigcapskip=-2pt 
	\subfigure[]{
		\includegraphics[width=70mm]{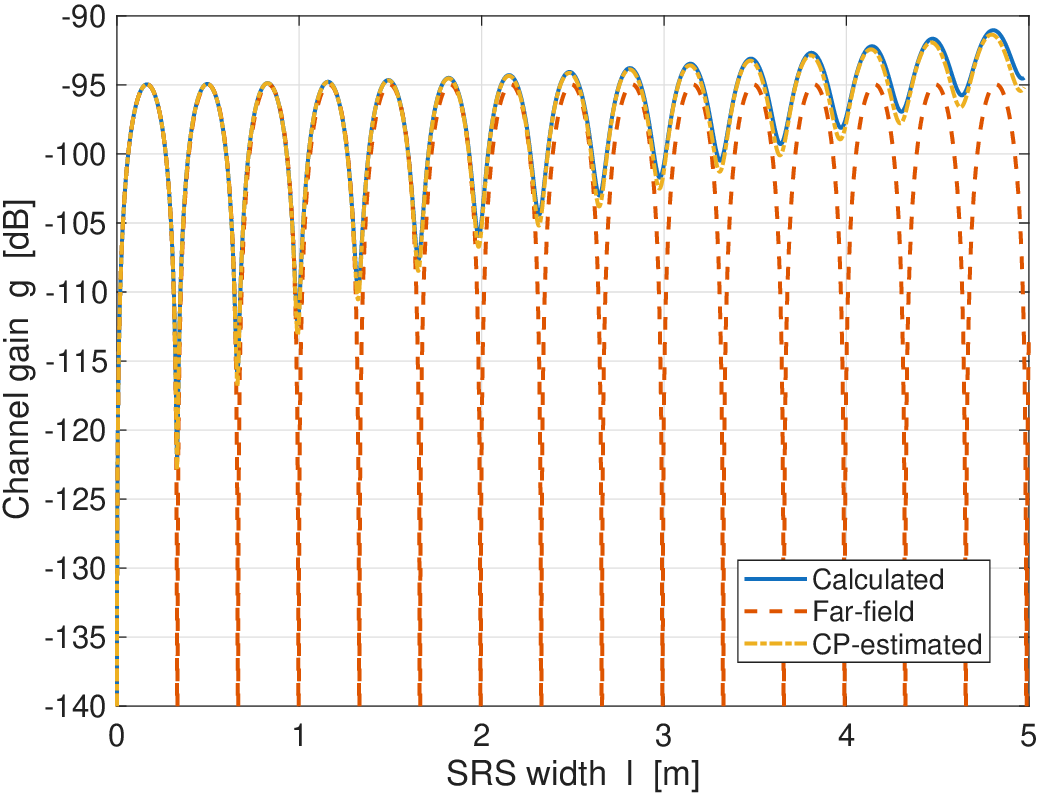} 
	}	\\	
	\subfigure[]{
		\includegraphics[width=70mm]{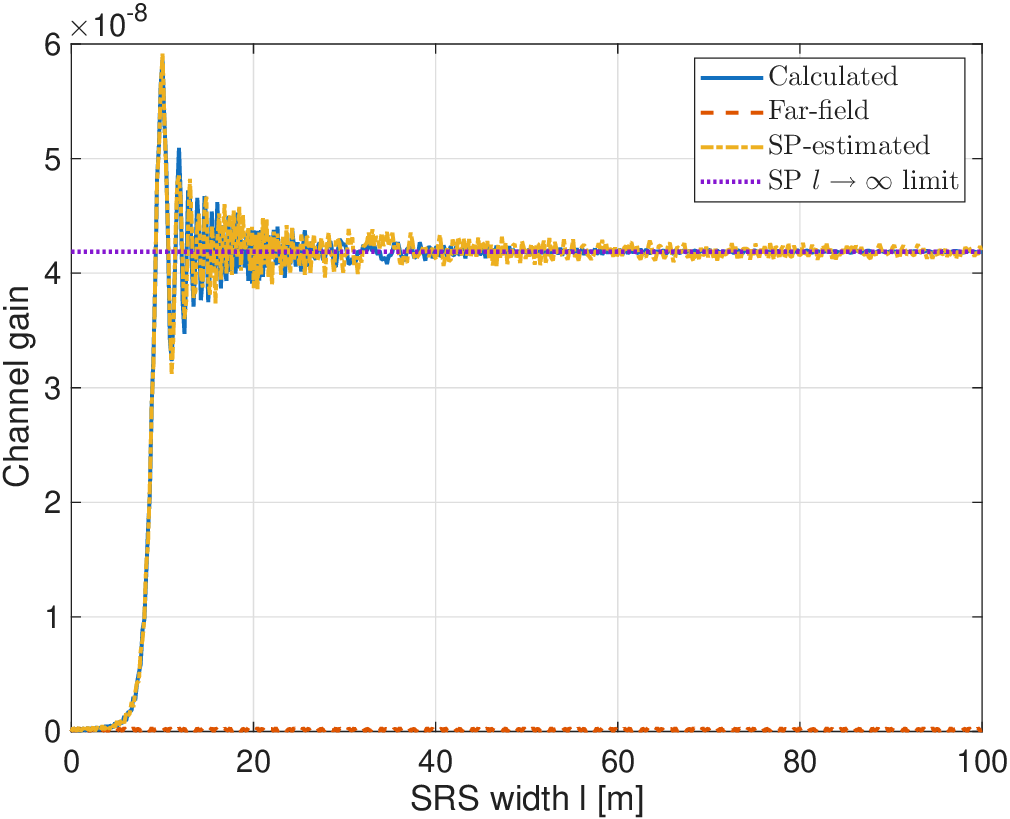} 
	}	
	\caption{Channel gain versus width of SRS $l$: (a) small aperture case, (b) large aperture case.}
	\label{fig_mis_Channel_gain_vs_width}
	\vspace{-4pt}
\end{figure}
We characterize the impact of the SRS aperture on the channel gain under the angle-misaligned configuration in Fig.~\ref{fig_mis_Channel_gain_vs_width}~(a). It can be observed from Fig.~\ref{fig_mis_Channel_gain_vs_width}~(a) that the channel gain is very small when the angles are misaligned, and it increases with mild oscillations as the aperture grows. Our analytical expression accurately captures both the overall increasing trend and the nonzero local minima. In contrast, the far-field expression exhibits noticeable errors around each local minimum and fails to reflect the increasing trend with the aperture. This is because, in practice, phase cancellation does not drive the cascaded-channel gain exactly to zero, whereas the far-field approximation incorrectly predicts perfect nulls and thus cannot capture the gradual gain increase.

We illustrate the asymptotic regime where the SRS aperture tends to infinity in Fig.~\ref{fig_mis_Channel_gain_vs_width}~(b). We observe from Fig.~\ref{fig_mis_Channel_gain_vs_width}~(b) that, even under angle misalignment, the channel gain first increases with the aperture, then decreases, and finally converges to a constant through damped oscillations. From the perspective of the intended user, this implies a wider angular coverage of non-negligible gain; meanwhile, it also indicates potentially stronger co-channel interference to other users. Moreover, the stationary-point-based closed-form expression remains accurate over a larger range of aperture sizes, although higher-order approximations are still required when the aperture becomes extremely large. In addition, the stationary-point-based asymptotic limit is highly accurate, whereas the far-field expression merely fluctuates around a small value that is far from the true gain.

\begin{figure}[htbp]
	\centering
	\includegraphics [width=70mm]{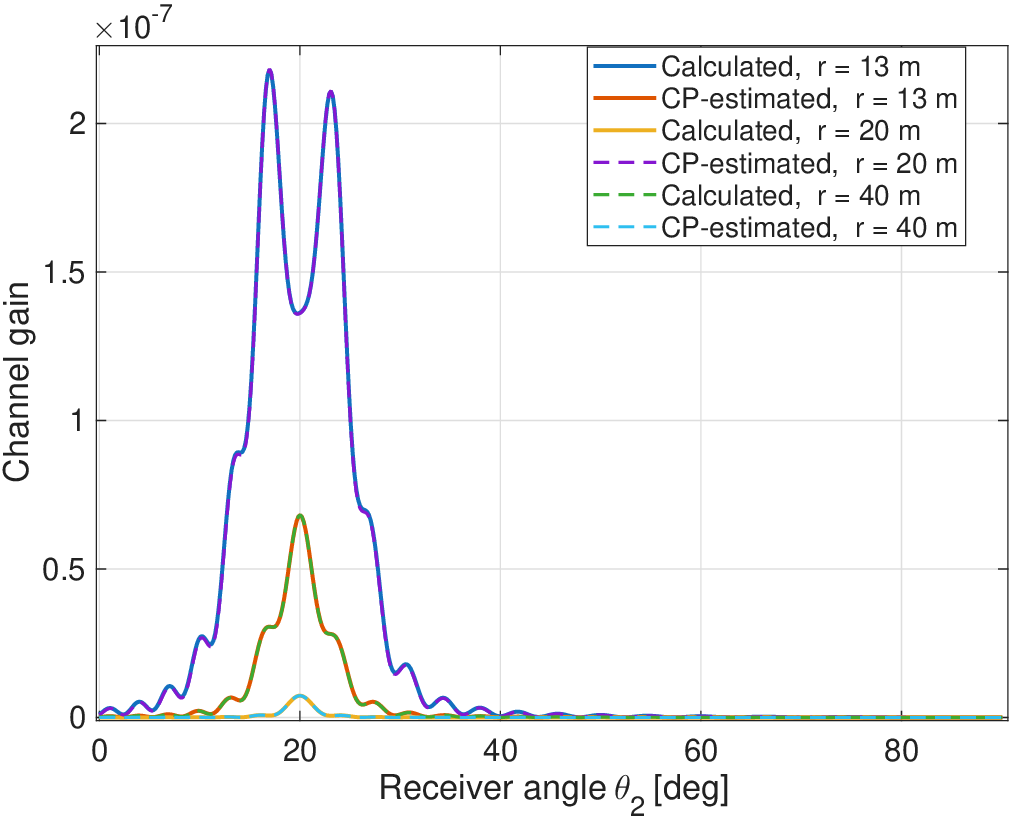}\\
	\caption{Channel gain vs reflected angle $\theta_{2}$.}
	\label{fig_Single_SRS_mis_angle}
\end{figure}
We examine the channel gain under different reflection angles in Fig.~\ref{fig_Single_SRS_mis_angle}. It can be observed from Fig.~\ref{fig_Single_SRS_mis_angle} that the closed-form expressions remain accurate across different angles. When the transceivers are closer to the SRS, the channel gain becomes higher and the angular region with non-negligible gain also expands. However, when the distance is relatively small, the maximum gain is no longer achieved at the specular-reflection angle; instead, the main beam shifts and splits into two lobes.

\begin{figure}[htbp]
	\centering
	\includegraphics [width=70mm]{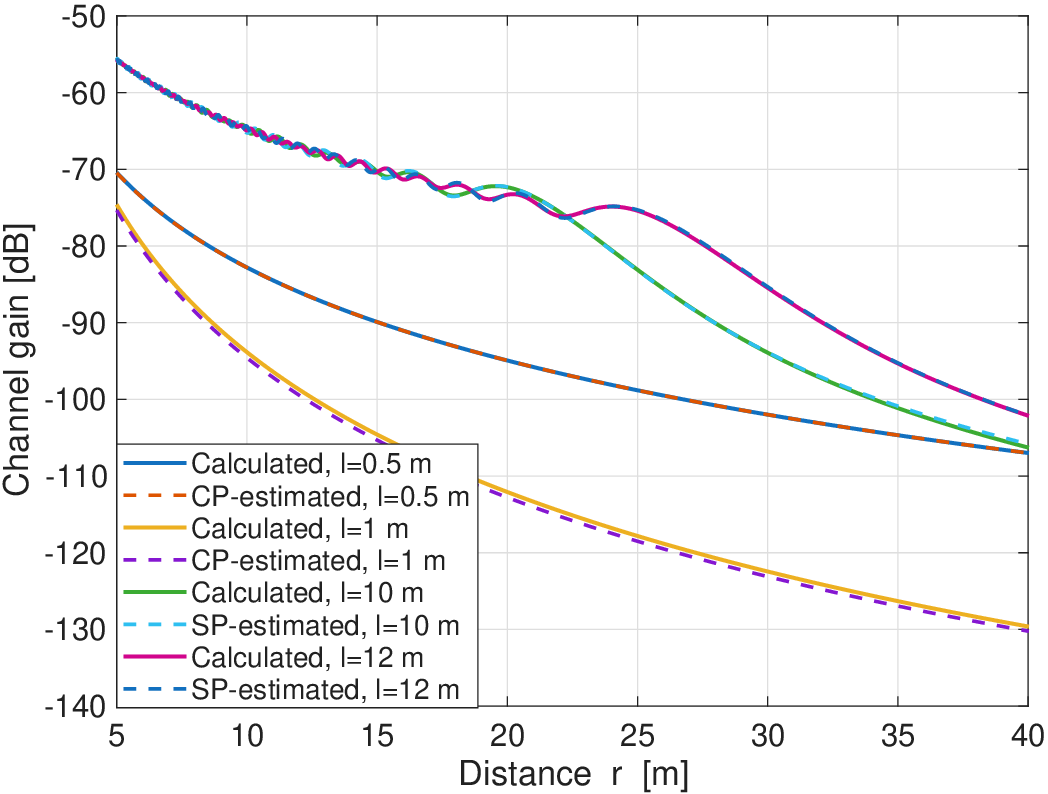}\\
	\caption{Channel gain vs distance.}
	\label{fig_Single_SRS_mis_distance}
\end{figure}
We show the impact of the transceiver distance on the angle-misaligned channel gain in Fig.~\ref{fig_Single_SRS_mis_distance}. We observe from Fig.~\ref{fig_Single_SRS_mis_distance} that, when the aperture is small, the center-point-based estimate is accurate and the channel gain decreases monotonically as the distance increases. When the aperture is large, the stationary-point-based estimate is accurate; as the distance increases, the gain decreases in an oscillatory manner at short ranges and then transitions to a monotonic decay at long ranges.

%
%

\begin{figure}[htbp]
	\centering
	\includegraphics [width=70mm]{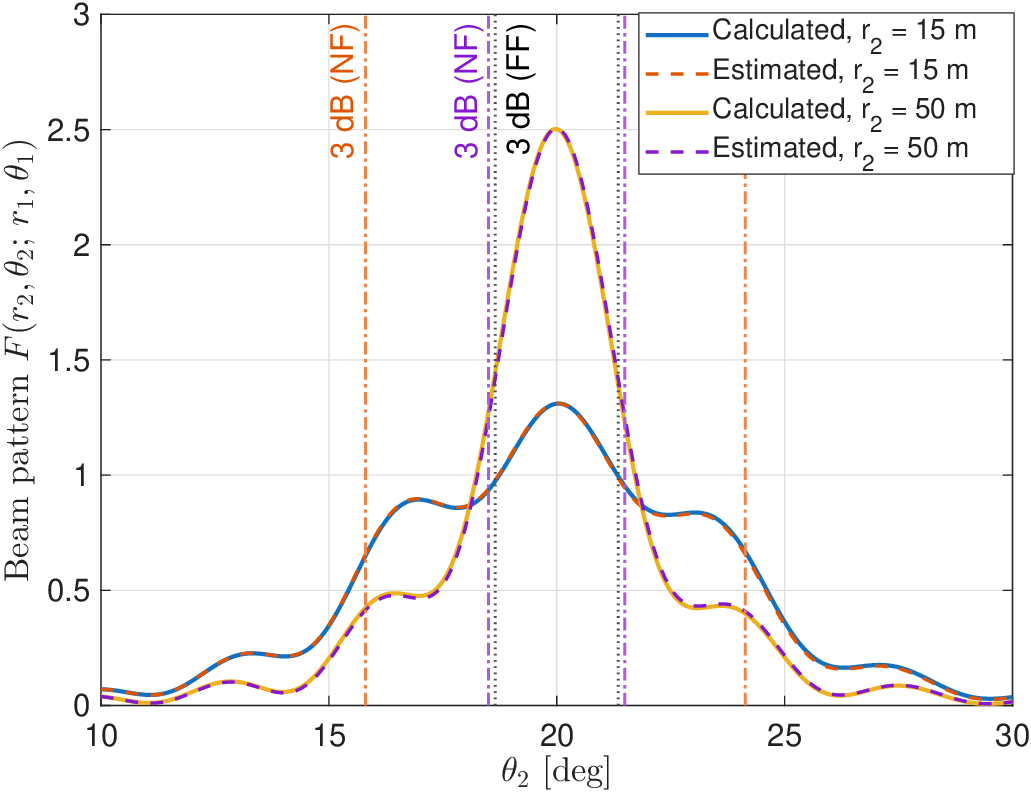}\\
	\caption{Angular reflected beam pattern.}
	\label{fig_beam_angular}
\end{figure}
We demonstrate the proposed beam pattern under different reflection angles in Fig.~\ref{fig_beam_angular}. Specifically, we calculate the exact beam pattern normalized by the path loss, and compare it with the estimation obtained from Eq.~(\ref{eq:41}). It can be observed from Fig.~\ref{fig_beam_angular} that the proposed approximation is accurate. Moreover, at long distances, the angular beamwidth is narrow, and the far-field 3-dB beamwidth closely matches the near-field 3-dB beamwidth. In contrast, at short distances, the angular beamwidth becomes wider, and only the near-field 3-dB beamwidth provides a valid characterization, while the far-field approximation breaks down.

\begin{figure}[htbp]
	\centering
	\includegraphics [width=70mm]{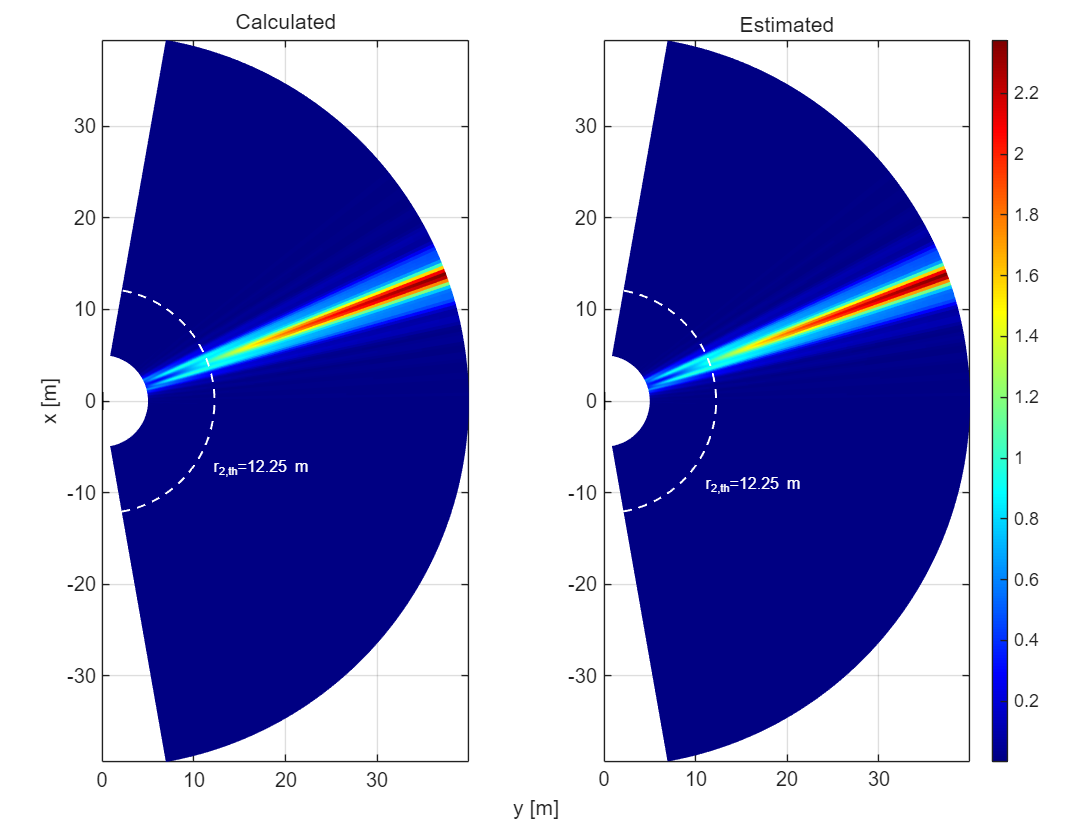}\\
	\caption{Reflected beam pattern in the planar view.}
	\label{fig_beam_pattern}
\end{figure}
We show the reflected beam pattern of SRS over the entire plane in Fig.~\ref{fig_beam_pattern}. It can be observed from Fig.~\ref{fig_beam_pattern} that our estimated beam pattern is accurate across the whole plane. Moreover, the reflected beam mainly affects only a small angular region centered around the specular-reflection direction. Although a single SRS cannot achieve near-field focusing, it exhibits a distance-dependent signal-combining effect: when the observation point is sufficiently far, the reflected components combine more constructively. In the near-field region, however, the main lobe may even vanish. Our derived threshold accurately characterizes the critical distance beyond which a main lobe exists.

\begin{figure}[htbp]
	\centering
	\includegraphics [width=70mm]{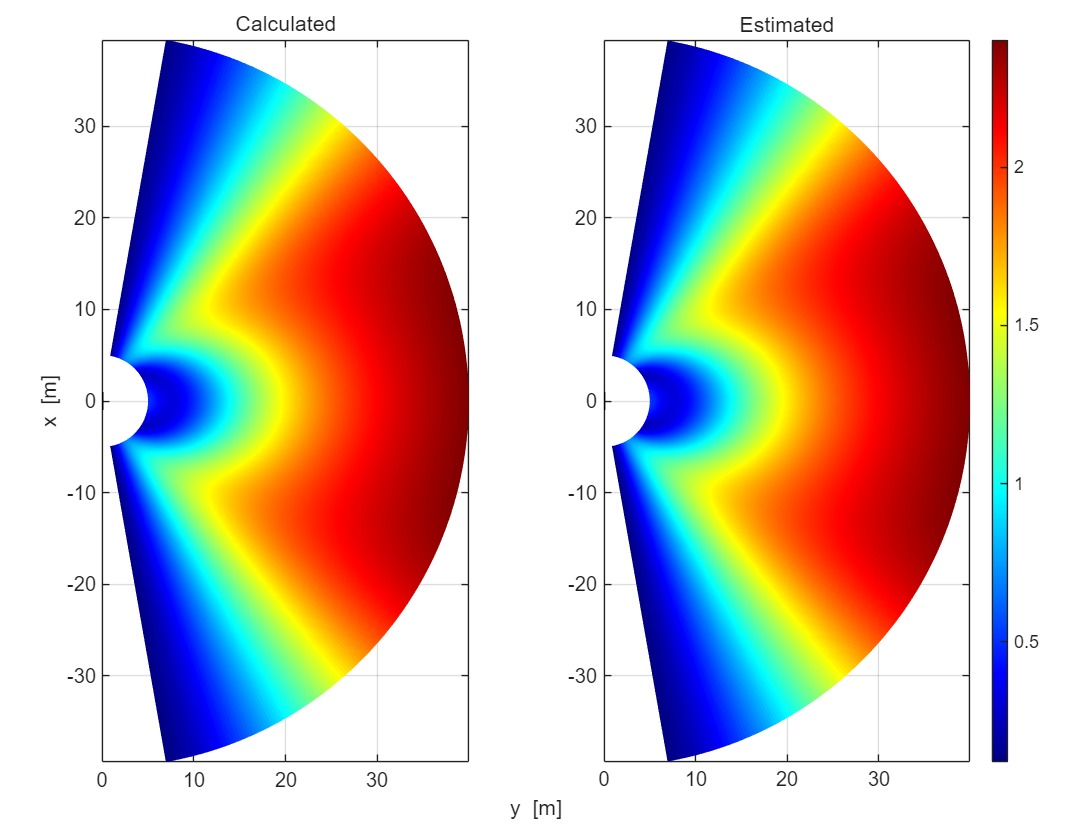}\\
	\caption{Effective reflecting area.}
	\label{fig_effective_eara}
\end{figure}
We set $r_1=20$~m and use the beam pattern in Eq.~(\ref{eq:41}) with the reflection angle equal to the incidence angle, i.e., the path-loss-normalized gain at each point under the angle-aligned configuration, to characterize the effective reflection region over the entire plane in Fig.~\ref{fig_effective_eara}. We observe from Fig.~\ref{fig_effective_eara} that, the effective region is constrained by both a minimum-distance threshold and a maximum-angle threshold. This is because longer distances are more favorable for constructive combining, while smaller reflection angles yield higher projected power. Note that this figure is obtained by sweeping the incidence and reflection angles with respect to a fixed (non-rotating) SRS. In a practical system, rotating the SRS can further enlarge the effective region over the plane.

\subsection{SRS-aided communication systems}
Finally, we investigate the performance of the SRS-aided communication systems. For the performance comparison, we consider the following benchmark schemes:
\begin{itemize}	
	\item{\bf Lower bound:} To highlight the necessity of rotation for achieving angle alignment, we take the performance of a non-rotating SRS as a lower bound.
	\item {\bf Upper bound:}  When the transmitter and receiver are located in the absolute far field of the SRS, all reflected signals add up constructively and the gain is maximized. In this case, the rate can be computed using the far-field approximation in Eq.~(\ref{eq:g_inifine_l}), and this rate can be regarded as a performance upper bound for the SRS.
	\item {\bf RIS:} For the RIS, the channel between the transmitter and the $n$-th RIS element (or between the $n$-th RIS element and the receiver) is modeled as $h_n = \sqrt{\alpha_n^{(\mathrm{RIS})}}\, e^{j\frac{2\pi}{\lambda} r_n}$, where $\alpha_n^{(\mathrm{RIS})} = \frac{A}{4\pi r_n^2} \cos\theta_n$, $r_n$ denotes the distance between them, $A$ is the effective width of each RIS element, and $\theta_n$ is the angle of incidence or reflection with respect to the element's surface normal. The optimal RIS beamforming can be expressed as $\phi_n = \exp(-j\frac{2\pi}{\lambda}(r_n^{(\mathrm{Tx-RIS})}+r_n^{(\mathrm{RIS-Rx})}))$. Typically, the effective aperture is set as $\frac{\lambda^2}{4\pi}$, corresponding to an effective width $A = \sqrt{\frac{\lambda^2}{4\pi}}$, and in practice RIS beamforming with $m$ discrete controllable phase levels is obtained by minimizing the Euclidean-distance.
	\item {\bf SRS array (SRS-A):} To better serve nearby users, the SRS can be partitioned into smaller SRS panels and assembled into an SRS array. Here, we keep the overall aperture size unchanged and, according to Eq.~(\ref{eq:optimal_psi}), steer each SRS-A element toward the user to preliminarily demonstrate the advantage of SRS-A. The achievable rate is computed as $R^{(\mathrm{SRS-A})}=\log_2\!\big(1+\frac{P_t}{\sigma^2}\big|\sum_{n} h_n(r_1^{(n)},r_2^{(n)},\frac{\theta_1^{(n)}+\theta_2^{(n)}}{2})\big|^2\big)$, where $n$ is the index of the SRS-A element. A more refined design is left for future work.
\end{itemize}

\begin{figure}[htbp]
	\centering
	\includegraphics [width=70mm]{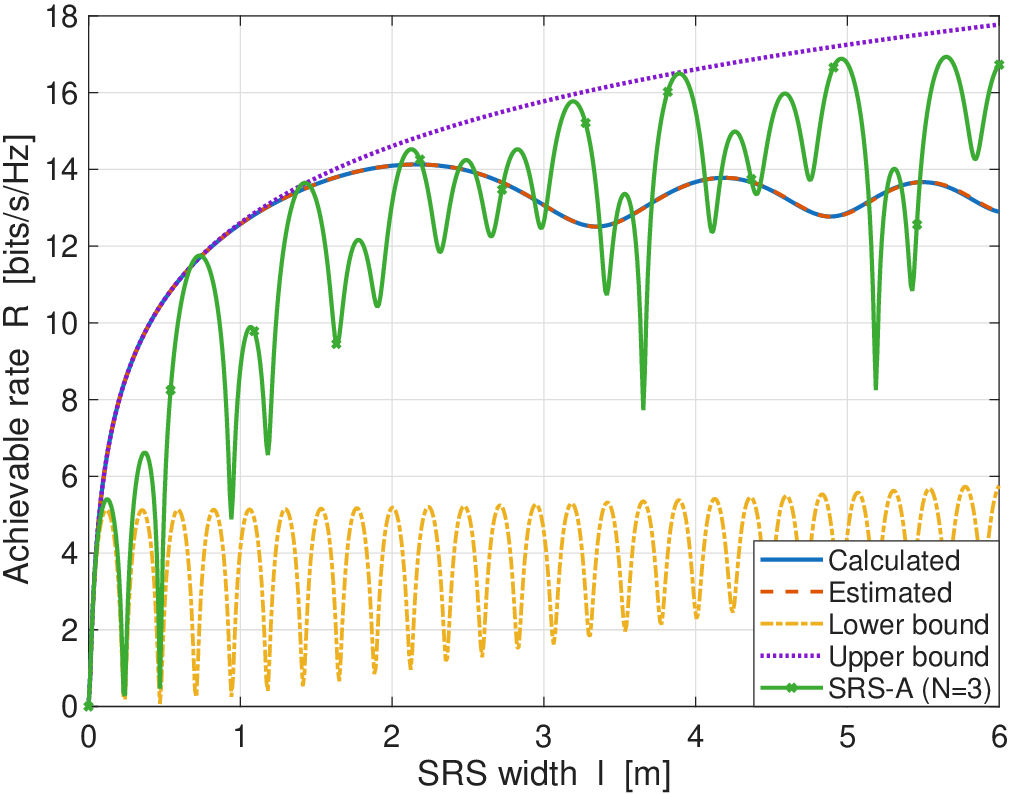}\\
	\caption{Achievable rate vs width of SRS $l$.}
	\label{fig_rate_l}
\end{figure}
We investigate the impact of the SRS width on the achievable rate In Fig.~\ref{fig_rate_l}. We set the incidence and reflection angles of the transmitter and receiver with respect to the SRS before rotation to $20^\circ$ and $50^\circ$, respectively. We observe from Fig.~\ref{fig_rate_l} that the analytical estimation is accurate. Moreover, rotating the SRS always yields a higher rate than the non-rotating case. When the user lies in the far-field region, the rotated SRS can essentially achieve the upper-bound performance. In the near field, however, a performance loss emerges because the reflected signals cannot be perfectly combined in phase. This limitation can be alleviated by employing an SRS array, which provides additional rotational degrees of freedom to better align the reflected components. In this example, each SRS element is aligned independently without joint optimization across elements. Hence, the performance of SRS-A is not yet optimal and exhibits noticeable fluctuations.

\begin{figure}[htbp]
	\centering
	\includegraphics [width=70mm]{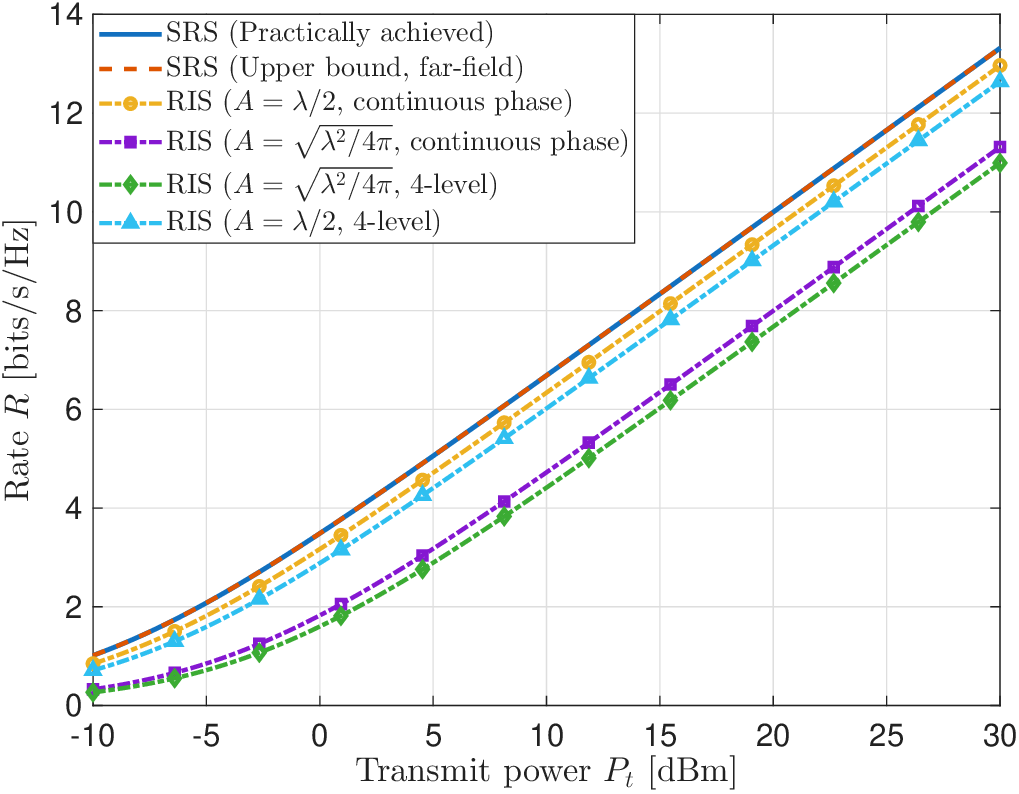}\\
	\caption{Achievable rate vs transmit power $P_t$.}
	\label{fig_rate_power}
\end{figure}
We illustrate the impact of the transmit power on the achievable rate in Fig.~\ref{fig_rate_power}. We set the SRS width to $0.5$~m, and the incidence/reflection angles of the transmitter and receiver before rotation to $20^\circ$ and $70^\circ$, respectively. We observe from Fig.~\ref{fig_rate_power} that, with a proper aperture size and distance, an SRS can achieve the far-field upper bound on the achievable rate. Furthermore, even an ideal RIS with a continuous aperture and continuous phase control yields a lower achievable rate than SRS. This is because the optimal power-projection direction coincides with the specular-reflection direction, and SRS can realize it via simple mechanical rotation, thereby achieving a higher power projection than RIS. Moreover, when considering a practically implementable RIS with the actual effective aperture and finite-resolution phase control, the performance gap between RIS and SRS becomes even larger. It should be noted that these results are obtained in free space, indicating that a single SRS may be more advantageous than an RIS with many elements when the objective is to establish a single strong LOS path. In practical scattering-rich environments, however, an RIS may better exploit NLOS components introduced by scatterers to achieve improved performance, albeit with substantially higher channel-estimation overhead than a single SRS.

\section{Conclusions}
In this paper, we have provided a comprehensive performance analysis for specular reflecting surface systems (SRS) in both the desired specular angle-aligned configuration and the general angle-misaligned case. The derived closed-form expressions reveal how the cascaded channel gain depends on the SRS aperture, carrier frequency, transceiver distances, and incidence angles. We have also investigated the far-field behavior and the asymptotic regime with an unbounded aperture, and characterized a channel-gain-based near-/far-field boundary. In addition, we have studied the SRS beam pattern and beamwidth, and effective operating region where a main lobe exists. Building on these analytical results, we have derived a closed-form achievable-rate expression for an SRS-aided communication system. Numerical results have validated the accuracy of the proposed expressions, highlighted the distinctive near-field behavior of SRS caused by non-constructive edge-aperture combining, and demonstrated that an SRS with a continuous aperture, continuous angular-resolution control, and stronger power projection can outperform RIS in the far-field regime.

\begin{appendices}
\end{appendices}

\bibliography{IEEEtran}

@ARTICLE{1137900,
	author={Sherman, J.},
	journal={IRE Transactions on Antennas and Propagation}, 
	title={Properties of focused apertures in the fresnel region}, 
	year={1962},
	volume={10},
	number={4},
	pages={399-408},
	doi={10.1109/TAP.1962.1137900}}

@book{mabramowitz64:handbook,
	address = {New York},
	editor = {Abramowitz, Milton and Stegun, Irene A.},
	publisher = {Dover Publications, Inc.},
	title = {Handbook of Mathematical Functions with Formulas, Graphs and Mathematical Tables},
	year = 1965
}

@ARTICLE{9866003,
	author={Zheng, Beixiong and Zhang, Rui},
	journal={IEEE Transactions on Wireless Communications}, 
	title={Simultaneous Transmit Diversity and Passive Beamforming With Large-Scale Intelligent Reflecting Surface}, 
	year={2023},
	volume={22},
	number={2},
	pages={920-933},
	doi={10.1109/TWC.2022.3199426}}

@ARTICLE{9184098,
	author={Björnson, Emil and Sanguinetti, Luca},
	journal={IEEE Open Journal of the Communications Society}, 
	title={Power Scaling Laws and Near-Field Behaviors of Massive {MIMO} and Intelligent Reflecting Surfaces}, 
	year={2020},
	volume={1},
	number={},
	pages={1306-1324},
	doi={10.1109/OJCOMS.2020.3020925}}

@ARTICLE{10697414,
	author={Fang, Yuguang and Deng, Yiqin and Chen, Xianhao},
	journal={IEEE Communications Magazine}, 
	title={Resources on the Move for Smart City: A Disruptive Perspective on the Grand Convergence of Sensing, Communications, Computing, Storage, and Intelligence}, 
	year={2025},
	volume={63},
	number={4},
	pages={200-206},
	doi={10.1109/MCOM.001.2400084}}

@ARTICLE{10559446,
	author={Huang, Qingxiao and Hu, Jie and Yang, Kun},
	journal={IEEE Transactions on Green Communications and Networking}, 
	title={Intelligent Reflecting Surface Assisted Integrated Data and Energy Multicast System in Terahertz-Bands}, 
	year={2025},
	volume={9},
	number={1},
	pages={152-163},
	doi={10.1109/TGCN.2024.3415030}}

@ARTICLE{10721321,
	author={Huang, Qingxiao and Hu, Jie and Zhao, Yizhe and Yang, Kun},
	journal={IEEE Transactions on Wireless Communications}, 
	title={Holographic Integrated Data and Energy Transfer}, 
	year={2024},
	volume={23},
	number={12},
	pages={18987-19002},
	doi={10.1109/TWC.2024.3477750}}

@ARTICLE{11212817,
	author={Huang, Qingxiao and Zhao, Yizhe and Hu, Jie and Yang, Kun and Fang, Yuguang},
	journal={IEEE Transactions on Wireless Communications}, 
	title={Circular Holographic {MIMO} Beamforming for Integrated Data and Energy Multicast Systems}, 
	year={2026},
	volume={25},
	number={},
	pages={5733-5748},
	doi={10.1109/TWC.2025.3620886}}

@ARTICLE{10470405,
	author={Huang, Qingxiao and Hu, Jie and Yue, Qingdong and Yang, Kun},
	journal={IEEE Wireless Communications Letters}, 
	title={Beamforming Design in Intelligent Reflecting Surface Aided Multicast System in Ultra-High Frequency Bands With Large-Scale Antenna Array}, 
	year={2024},
	volume={13},
	number={5},
	pages={1429-1433},
	doi={10.1109/LWC.2024.3373494}}

@ARTICLE{4066064,
	author={Ryerson, J. L.},
	journal={Proceedings of the IRE}, 
	title={Passive Satellite Communication}, 
	year={1960},
	volume={48},
	number={4},
	pages={613-619},
	doi={10.1109/JRPROC.1960.287436}}

@article{Jakes1961EchoI,
	author  = {Jakes, W.},
	title   = {A Transatlantic Communication Experiment via Echo I Satellite},
	journal = {Nature},
	volume  = {190},
	pages   = {709},
	year    = {1961},
	doi     = {10.1038/190709a0},
	url     = {https://doi.org/10.1038/190709a0}
}

@ARTICLE{1687092,
	author={Reber, Grote},
	journal={Proceedings of the IRE}, 
	title={Cosmic Static}, 
	year={1940},
	volume={28},
	number={2},
	pages={68-70},
	doi={10.1109/JRPROC.1940.228921}}

@INPROCEEDINGS{9013979,
	author={Zhang, Lan and Chen, Xianhao and Fang, Yuguang and Huang, Xiaoxia and Fang, Xuming},
	booktitle={2019 IEEE Global Communications Conference {(GLOBECOM)}}, 
	title={Learning-Based mmWave V2I Environment Augmentation through Tunable Reflectors}, 
	year={2019},
	volume={},
	number={},
	pages={1-6},
	doi={10.1109/GLOBECOM38437.2019.9013979}}

@ARTICLE{9082859,
	author={Zhang, Lan and Yan, Li and Lin, Bin and Ding, Haichuan and Fang, Yuguang and Fang, Xuming},
	journal={IEEE Transactions on Vehicular Technology}, 
	title={Augmenting Transmission Environments for Better Communications: Tunable Reflector Assisted {MmWave} WLANs}, 
	year={2020},
	volume={69},
	number={7},
	pages={7416-7428},
	doi={10.1109/TVT.2020.2991547}}

@ARTICLE{7914640,
	author={Xue, Qing and Fang, Xuming and Wang, Cheng-Xiang},
	journal={IEEE Journal on Selected Areas in Communications}, 
	title={Beamspace {SU-MIMO} for Future Millimeter Wave Wireless Communications}, 
	year={2017},
	volume={35},
	number={7},
	pages={1564-1575},
	doi={10.1109/JSAC.2017.2699085}}

@inproceedings{10.1145/2342356.2342440,
	author = {Zhou, Xia and Zhang, Zengbin and Zhu, Yibo and Li, Yubo and Kumar, Saipriya and Vahdat, Amin and Zhao, Ben Y. and Zheng, Haitao},
	title = {Mirror mirror on the ceiling: flexible wireless links for data centers},
	year = {2012},
	isbn = {9781450314190},
	publisher = {Association for Computing Machinery},
	address = {New York, NY, USA},
	url = {https://doi.org/10.1145/2342356.2342440},
	doi = {10.1145/2342356.2342440},
	booktitle = {Proceedings of the ACM SIGCOMM 2012 Conference on Applications, Technologies, Architectures, and Protocols for Computer Communication},
	pages = {443–454},
	numpages = {12},
	location = {Helsinki, Finland},
	series = {SIGCOMM '12}
}

@ARTICLE{9384308,
	author={Li, Songlin and Guo, Shisheng and Chen, Jiahui and Yang, Xiaqing and Fan, Shihao and Jia, Chao and Cui, Guolong and Yang, Haining},
	journal={IEEE Transactions on Vehicular Technology}, 
	title={Multiple Targets Localization Behind L-Shaped Corner via {UWB} Radar}, 
	year={2021},
	volume={70},
	number={4},
	pages={3087-3100},
	doi={10.1109/TVT.2021.3068266}}

@ARTICLE{9374714,
	author={Anjinappa, Chethan Kumar and Erden, Fatih and Güvenç, Ismail},
	journal={IEEE Transactions on Vehicular Technology}, 
	title={Base Station and Passive Reflectors Placement for Urban {mmWave} Networks}, 
	year={2021},
	volume={70},
	number={4},
	pages={3525-3539},
	doi={10.1109/TVT.2021.3065221}}

@ARTICLE{11007274,
	author={Ning, Boyu and Yang, Songjie and Wu, Yafei and Wang, Peilan and Mei, Weidong and Yuen, Chau and Björnson, Emil},
	journal={IEEE Wireless Communications}, 
	title={Movable Antenna-Enhanced Wireless Communications: General Architectures and Implementation Methods}, 
	year={2025},
	volume={32},
	number={5},
	pages={108-116},
	doi={10.1109/MWC.013.2400238}}

@misc{zheng2025,
	title={Rotatable Antenna Enabled Wireless Communication: Modeling and Optimization}, 
	author={Beixiong Zheng and Qingjie Wu and Tiantian Ma and Rui Zhang},
	year={2025},
	eprint={2501.02595},
	archivePrefix={arXiv},
	primaryClass={cs.IT},
	url={https://arxiv.org/abs/2501.02595}, 
}

@ARTICLE{8910627,
	author={Wu, Qingqing and Zhang, Rui},
	journal={IEEE Communications Magazine}, 
	title={Towards Smart and Reconfigurable Environment: Intelligent Reflecting Surface Aided Wireless Network}, 
	year={2020},
	volume={58},
	number={1},
	pages={106-112},
	doi={10.1109/MCOM.001.1900107}}

@ARTICLE{9140329,
	author={Di Renzo, Marco and Zappone, Alessio and Debbah, Merouane and Alouini, Mohamed-Slim and Yuen, Chau and de Rosny, Julien and Tretyakov, Sergei},
	journal={IEEE Journal on Selected Areas in Communications}, 
	title={Smart Radio Environments Empowered by Reconfigurable Intelligent Surfaces: How It Works, State of Research, and The Road Ahead}, 
	year={2020},
	volume={38},
	number={11},
	pages={2450-2525},
	doi={10.1109/JSAC.2020.3007211}}

@ARTICLE{10850658,
	author={An, Jiancheng and Yuen, Chau and Renzo, Marco Di and Debbah, Mérouane and Poor, H. Vincent and Hanzo, Lajos},
	journal={IEEE Transactions on Wireless Communications}, 
	title={Flexible Intelligent Metasurfaces for Downlink Multiuser {MISO} Communications}, 
	year={2025},
	volume={24},
	number={4},
	pages={2940-2955},
	doi={10.1109/TWC.2025.3526843}}

@ARTICLE{10558818,
	author={An, Jiancheng and Yuen, Chau and Dai, Linglong and Di Renzo, Marco and Debbah, Mérouane and Hanzo, Lajos},
	journal={IEEE Wireless Communications}, 
	title={Near-Field Communications: Research Advances, Potential, and Challenges}, 
	year={2024},
	volume={31},
	number={3},
	pages={100-107},
	doi={10.1109/MWC.004.2300450}}

@ARTICLE{9136592,
	author={Huang, Chongwen and Hu, Sha and Alexandropoulos, George C. and Zappone, Alessio and Yuen, Chau and Zhang, Rui and Renzo, Marco Di and Debbah, Merouane},
	journal={IEEE Wireless Communications}, 
	title={Holographic {MIMO} Surfaces for 6G Wireless Networks: Opportunities, Challenges, and Trends}, 
	year={2020},
	volume={27},
	number={5},
	pages={118-125},
	doi={10.1109/MWC.001.1900534}}

@book{zhang2021reconfigurable,
	title={Reconfigurable intelligent surface-empowered {6G}},
	author={Zhang, Hongliang and Di, Boya and Song, Lingyang and Han, Zhu},
	year={2021},
	publisher={Springer}
}

@article{Liang2021RISSmartWirelessEnvironments,
	author  = {Liang, Ying-Chang and Chen, Jie and Long, Ruizhe and He, Zhen-Qing and Lin, Xianqi and Huang, Chenlu and Liu, Shilin and Shen, Xuemin (Sherman) and Di Renzo, Marco},
	title   = {Reconfigurable Intelligent Surfaces for Smart Wireless Environments: Channel Estimation, System Design and Applications in {6G} Networks},
	journal = {Science China Information Sciences},
	volume  = {64},
	number  = {10},
	pages   = {200301:1--200301:21},
	year    = {2021},
	doi     = {10.1007/s11432-020-3261-5},
	url     = {https://doi.org/10.1007/s11432-020-3261-5}
}

@ARTICLE{10989638,
	author={Lu, Haiquan and Yu, Zhi and Zeng, Yong and Ma, Shaodan and Jin, Shi and Zhang, Rui},
	journal={IEEE Transactions on Wireless Communications}, 
	title={Wireless Communication With Flexible Reflector: Joint Placement and Rotation Optimization for Coverage Enhancement}, 
	year={2025},
	volume={24},
	number={10},
	pages={8252-8266},
	doi={10.1109/TWC.2025.3564956}}

@INPROCEEDINGS{10279522,
	author={Yu, Zhi and Feng, Chao and Zeng, Yong and Li, Teng and Jin, Shi},
	booktitle={ICC 2023 - IEEE International Conference on Communications}, 
	title={Wireless Communication Using Metal Reflectors: Reflection Modelling and Experimental Verification}, 
	year={2023},
	volume={},
	number={},
	pages={4701-4706},
	doi={10.1109/ICC45041.2023.10279522}}

@misc{zheng2025sensing,
	title={Wireless Sensing with Movable Intelligent Surface}, 
	author={Ziyuan Zheng and Qingqing Wu and Yanze Zhu and Wen Chen and Ying Gao and Honghao Wang},
	year={2025},
	eprint={2509.15627},
	archivePrefix={arXiv},
	primaryClass={eess.SP},
	url={https://arxiv.org/abs/2509.15627}, 
}

@ARTICLE{11082321,
	author={Xu, Bohui and Lin, Bangjiang and Chen, Jian and Zheng, Bowen and Pang, Guojun and Luo, Jiabin and Ghassemlooy, Zabih},
	journal={IEEE Internet of Things Journal}, 
	title={{OIRS}-Assisted {NLOS} Visible Light Communication Systems: Modeling, Optimization, and Experimental Validation}, 
	year={2025},
	volume={12},
	number={19},
	pages={40458-40469},
	doi={10.1109/JIOT.2025.3589633}}

@article{ma2024multi,
	title={Multi-hop multi-{RIS} wireless communication systems: Multi-reflection path scheduling and beamforming},
	author={Ma, Xiaoyan and Zhang, Haixia and Chen, Xianhao and Fang, Yuguang and Yuan, Dongfeng},
	journal={IEEE Transactions on Wireless Communications},
	volume={23},
	number={7},
	pages={6778--6792},
	year={2024},
	publisher={IEEE}
}

@InProceedings{zhang2019tunable,
	author =        {L. Zhang and L. Yan and B. Lin and Y. Fang and X. Fang},
	title =         {Tunable reflectors enabled environment augmentation for better {mmWave WLANs}},
	booktitle =     {IEEE/CIC International Conference on Communications in China (ICCC)},
	year =      {2019},
	address =   {Changchun, Jilin, China},
	month =     {August 11-13}
}

@misc{ren2026,
	title={Shatter Throughput Ceilings: Leveraging Reflection Surfaces to Enhance Transmissions for Vehicular Fast Data Exchange}, 
	author={Qianyao Ren and Qingxiao Huang and Yiqin Deng and Xianhao Chen and Phone Lin and Yuguang Fang},
	year={2026},
	eprint={2603.02752},
	archivePrefix={arXiv},
	primaryClass={cs.IT},
	url={https://arxiv.org/abs/2603.02752}, 
}

@ARTICLE{11433651,
	author={Zhao, Yizhe and Zhang, Long and Yang, Halvin and Yang, Kun and Zhang, Rui and Song, Lingyang and Liu, Yuanwei},
	journal={IEEE Communications Surveys \& Tutorials}, 
	title={Reconfigurable Antennas for Next-Generation Mobile Communication Networks: A Comprehensive Survey and Tutorial}, 
	year={2026},
	volume={28},
	number={},
	pages={5267-5306},
	doi={10.1109/COMST.2026.3673688}}
\bibliographystyle{ieeetr}

\end{document}